\documentclass[sigconf]{acmart}
\usepackage{multirow}
\usepackage{colortbl}
\usepackage{amsmath}
\usepackage{enumitem}
\usepackage{algorithm}
\usepackage{algpseudocode}
\usepackage{float}
\usepackage{graphicx}
\usepackage{array}
\usepackage{subcaption}
\usepackage[normalem]{ulem}
\useunder{\uline}{\ul}{}
\algnewcommand\algorithmicforeach{\textbf{for each}}
\algdef{S}[FOR]{ForEach}[1]{\algorithmicforeach\ #1\ \algorithmicdo}
\newcolumntype{V}[1]{>{\centering\arraybackslash\rotatebox{90}}m{#1}}

\AtBeginDocument{%
  }

\copyrightyear{2025}
\acmYear{2025}
\setcopyright{rightsretained}
\acmConference[SIGIR '25] {Proceedings of the 48th International ACM SIGIR Conference on Research and Development in Information Retrieval}{ July 13--18, 2025}{Padua, Italy.}
\acmBooktitle{Proceedings of the 48th International ACM SIGIR Conference on Research and Development in Information Retrieval (SIGIR '25), July 13--18, 2025, Padua, Italy}
\acmISBN{979-8-4007-1592-1/25/07}
\acmDOI{10.1145/XXXXXX.XXXXXX}

\begin{document}

\title{Diversity-aware Dual-promotion Poisoning Attack on\\ Sequential Recommendation}

\author{Yuchuan Zhao}
\affiliation{%
  \institution{The University of Queensland}
  \city{Brisbane}
  \state{Queensland}
  \country{Australia}
}
\email{y.c.zhao@uq.edu.au}

\author{Tong Chen}
\affiliation{%
  \institution{The University of Queensland}
  \city{Brisbane}
  \state{Queensland}
  \country{Australia}
}
\email{tong.chen@uq.edu.au}

\author{Junliang Yu}
\affiliation{%
  \institution{The University of Queensland}
  \city{Brisbane}
  \state{Queensland}
  \country{Australia}
}
\email{jl.yu@uq.edu.au}

\author{Kai Zheng}
\affiliation{%
 \institution{University of Electronic Science and Technology of China}
 \city{Chengdu}
 \state{Sichuan}
 \country{China}}
\email{zhengkai@uestc.edu.cn}

\author{Lizhen Cui}
\authornote{Hongzhi Yin and Lizhen Cui are co-corresponding authors.}
\affiliation{%
  \institution{Shandong University}
  \city{Jinan}
  \state{Shandong}
  \country{China}}
\email{clz@sdu.edu.cn}

\author{Hongzhi Yin}
\authornotemark[1]
\affiliation{%
  \institution{The University of Queensland}
  \city{Brisbane}
  \state{Queensland}
  \country{Australia}
}
\email{h.yin1@uq.edu.au}

\renewcommand{\shortauthors}{Yuchuan Zhao et al.}

\begin{abstract}
Sequential recommender systems (SRSs) excel in capturing users' dynamic interests, thus playing a key role in various industrial applications. The popularity of SRSs has also driven emerging research on their security aspects, where data poisoning attack for targeted item promotion is a typical example. Existing attack mechanisms primarily focus on increasing the ranks of target items in the recommendation list by injecting carefully crafted interactions (i.e., poisoning sequences), which comes at the cost of demoting users' real preferences. Consequently, noticeable recommendation accuracy drops are observed, restricting the stealthiness of the attack. Additionally, the generated poisoning sequences are prone to substantial repetition of target items, which is a result of the unitary objective of boosting their overall exposure and lack of effective diversity regularizations. Such homogeneity not only compromises the authenticity of these sequences, but also limits the attack effectiveness, as it ignores the opportunity to establish sequential dependencies between the target and many more items in the SRS. To address the issues outlined, we propose a \textbf{D}iversity-aware \textbf{D}ual-promotion \textbf{S}equential \textbf{P}oisoning attack method named \(\textbf{DDSP}\) for SRSs. Specifically, by theoretically revealing the conflict between recommendation and existing attack objectives, we design a revamped attack objective that promotes the target item while maintaining the relevance of preferred items in a user's ranking list. We further develop a diversity-aware, auto-regressive poisoning sequence generator, where a re-ranking method is in place to sequentially pick the optimal items by integrating diversity constraints. By attacking two representative SRSs on three real-world datasets, comprehensive experimental results demonstrate that DDSP outperforms state-of-the-art attack methods in attack effectiveness. Moreover, DDSP achieves the strongest stealthiness with its lowest impact on recommendation accuracy.
\end{abstract}



\maketitle

\section{Introduction}
Sequential Recommender Systems (SRSs) \cite{fang2020deep,chen2022intent} are designed to capture user’s evolving interests and the dynamic nature of item-to-item transition patterns. In recent years, numerous neural-network-based models (e.g., SASRec \cite{kang2018self}, CL4SRec \cite{xie2022contrastive}) have emerged and demonstrated promising performance in sequential recommendation tasks. However, the openness of the real-world Recommender Systems (RSs), combined with the potential benefits of exploiting them, creates opportunities and incentives for malicious actors to launch attacks \cite{lam2004shilling, huang2025trustworthiness, long2024physical}. Recent studies have revealed that recommendation models are particularly susceptible to data poisoning attacks \cite{zheng2024poisoning,zhang2022pipattack,wu2024accelerating}, where attackers can inject carefully crafted data into the system to steer the recommendations in their favor. Among these, promotion attacks \cite{fang2018poisoning,chen2024adversarial,lin2020attacking} are especially prevalent. Through tampering with reviews and ratings \cite{nguyen2024manipulating}, merchants can promote specific items (i.e. target items) to a wide user base, thereby artificially boosting sales. To bolster the robustness and security of various recommendation services, studying and addressing these attacks is essential, and this is no exception for SRSs.

Existing poisoning attacks generally can be categorized into two types \cite{wang2024unveiling}, distinguished by the approaches employed to construct the poisoning data. The first type is heuristic-based methods which rely on manually designed rules. For instance, attackers may fabricate item co-occurrences with popular items under the assumption that frequently co-selected items are highly correlated \cite{yang2017fake}. However, these training-free methods often fail to capture complex user behaviors and model dynamics, leading to suboptimal attack performance. In contrast, optimization-based attacks explicitly maximize the attack objectives to forge user interactions \cite{fang2020influence,li2016data} or to tune the parameters of neural networks to generate optimal fake user profiles \cite{lin2020attacking,lin2022shilling}. A typical optimization strategy is to adopt ``bi-level optimization'' \cite{tang2020revisiting}, where a surrogate model \cite{yoon2024debiased} stands in for the victim recommendation model, and is iteratively updated to produce interactions that align with the attacker's goals. While this paradigm has demonstrated notable effectiveness, it still exhibits two major limitations:

\begin{figure}
    \setlength{\abovecaptionskip}{0.0cm}
    \setlength{\belowcaptionskip}{0.0cm}
    \centering
    \includegraphics[width=0.7\linewidth]{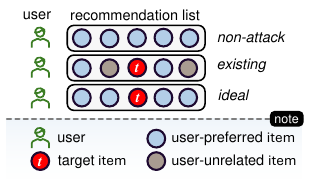}
    \caption{A toy example illustrating the comparison of a user’s recommendation lists across three scenarios: non-attack, existing attack, and ideal attack (proposed method).}
    \label{intro-case}
\end{figure}

\begin{itemize}[leftmargin=*]
    \item \textbf{Compromised attack stealthiness}. Most bi-level optimization-based attack objectives focus on improving the ranks of target items at the expense of demoting users' real preferences \cite{li2016data, tang2020revisiting, zhang2021data}. As a consequence, the recommendation accuracy of the victim model may be conspicuously degraded, undermining the stealthiness of the attack. Fig. \ref{intro-case} illustrates this issue intuitively: in the ``\textit{non-attack}'' scenario, the recommendation model is trained solely on the recommendation objective, positioning the user closer to the ground-truth item. However, in the``\textit{existing attack}'' scenario, the objective prioritizes the target item's ranking, resulting in the demotion of user preferred items and the inclusion of unrelated items in the recommendation list. This noticeable degradation in accuracy increases the likelihood of the attack being detected. 
    \item \textbf{Item repetition in generated sequences}. When generating malicious sequences against SRSs for item promotion, a common strategy is to adopt auto-regressive \cite{wang2023poisoning, yue2021black} with the well-trained surrogate model. However, during the bi-level model training, the surrogate model often overemphasizes the target items, leading to their repeated inclusion in the generated sequences \cite{wei2021model}. This repetition not only undermines the authenticity of the sequences, increasing the risk of being detected, but also reduces the likelihood of target items co-occurring with other items, diminishing the overall effectiveness of the attack.
\end{itemize}

To tackle these limitations, we propose a novel practical \textbf{D}iversity-aware \textbf{D}ual-promotion \textbf{S}equential \textbf{P}oisoning attack, named \textbf{DDSP}. For the first limitation, we introduce a \textit{dual-promotion attack objective} that avoids over-prioritizing target items during the surrogate model training. This approach allows both the target item and the ground-truth item to be drawn closer to the user in the feature space. As depicted in Fig. \ref{intro-case}, DDSP can achieve the ``\textit{ideal}'' scenario which simultaneously promotes both items without significantly degrading recommendation accuracy. Furthermore, by incorporating a contrastive regularization term, the target item is not only aligned with the user’s preferred items but also pushed away from unrelated items. This enhances the likelihood of effectively recommending both the target item and the user preferred items while preserving the stealthiness of the attack, reducing the risk of detection.

To overcome the second limitation, we propose a \textit{diversity-aware sequence generation} strategy that uses a re-ranking method to increase the diversity of generated sequences. Specifically, we generate a set of candidate next items in an auto-regressive manner and use two search strategies to measure the pairwise diversity of items within the sequence. Items are sequentially selected from the candidate set based on a comprehensive consideration of both relevance and diversity scores, continuing until the sequence reaches its maximum length. By introducing diversity, our method reduces the likelihood of item duplication and enhances the authenticity of the generated sequences. Moreover, this approach increases the co-occurrence rate of the target item with a broader range of items, thereby improving the success rate of the attack. To summarize, we have the following main contributions: 

\begin{itemize}[leftmargin=*]
    \item We propose a \textit{dual-promotion attack objective} that enhances the attacker's effectiveness while maintaining its stealthiness by simultaneously promoting both the target items and the user’s preferred items.
    \item  We introduce a \textit{diversity-aware sequence generation} strategy with re-ranking to address the issue of ``item repetition'', effectively enhancing the diversity and authenticity of the generated poisoning sequences.
    \item We evaluate DDSP against two representative sequential recommendation models on three real-world datasets, demonstrating its advantages in both stealthiness and attack efficacy.
\end{itemize}

\section{Related Works}
\textbf{Sequential Recommendation.} SRSs aim to predict users' next action by modeling users’ temporal and sequential interaction patterns \cite{kang2018self}. Early work often relies on markov decision processes \cite{rendle2010factorizing,he2016fusing}. With the advent of deep learning, RNNs \cite{cho2014learning} and their variants, such as LSTM \cite{graves2012long} and GRU \cite{donkers2017sequential}, become widely used for capturing intricate user interactions and item-to-item transitions. Beyond RNNs, more advanced neural architectures have emerged to capture user behavior characteristics; for instance, SASRec \cite{kang2018self} leverages self-attention \cite{vaswani2017attention} to predict user actions. Recently, significant progress has been made in incorporating self-supervised learning into the training of SRSs \cite{chen2022intent,xie2022contrastive,zhou2020s3}, with approaches like CL4SRec \cite{xie2022contrastive} harnessing contrastive learning \cite{wu2018unsupervised} to derive self-supervision signals from the observed user behavior. Despite these advancements, SRS models remain susceptible to malicious attacks, presenting a persistent and critical challenge.

\noindent\textbf{Data Poisoning Attacks on Recommender Systems (RSs).} Early poisoning attacks are often heuristic. For example, the random attack \cite{lin2022shilling} simply injects items randomly, whereas the bandwagon attack \cite{o2005recommender} interacts with popular items to increase the target item's visibility. These methods usually yield suboptimal performance \cite{li2022black} due to their static and non-optimizable nature. Subsequent work shifts towards optimization-based attacks. For instance, PGA \cite{li2016data} and \textit{S}-attack \cite{fang2020influence} optimize poisoning models for matrix-factorization-based RSs. As deep learning (DL) gains traction, attackers begin crafting fake user profiles via well-trained DL-based attack models \cite{huang2021data, wang2023revisiting, wang2024unveiling}. 

In SRSs, attackers must consider item dependencies. Prior work like LOKI \cite{zhang2020practical} employs RL to generate adversarial sequences but rely on impractical full access to user interactions. Yue et al. \cite{yue2021black} propose an extraction attack method that distills a surrogate model via repeated queries, increasing detection risk \cite{stevanovic2012feature}. Wang et al. \cite{wang2023poisoning} adopt a GAN-based method to create fake sequences by designating the target item as the label, ensuring its inclusion. However, none of these works take diversity into account during sequence generation, limiting their realism and stealth.

\section{Preliminaries}
\subsection{Sequential Recommendation Task}\label{sec:SRSs}
An SRS comprises a set of users \(\mathcal{U}\) and items \(\mathcal{I}\), where \(u \in \mathcal{U}\) represents a user, and \(i \in \mathcal{I}\) represents an item. 
Each user \(u\)'s interaction sequence is defined as \(s_u = \{i^{(u)}_{m}\}_{m=1}^M\), where \(i^{(u)}_m\) (\(1 \leq m \leq M\)) represents the \(m\)-th item interacted by user \(u\), arranged chronologically. \(M\) is the maximum sequence length. The set of interaction sequence for all users is \(\mathcal{S}\), with $|\mathcal{S}|=N$, where $N$ is the number of users. The task of sequential recommendation aims to predict the item that user \(u\) is most likely to interact with at the subsequent time step \(m+1\), based on the historical interaction. It can be formalized as estimating the probability of each item for user \(u\) at time step \(m+1\):
\begin{equation}
\setlength{\abovecaptionskip}{0.0cm}
\setlength{\belowcaptionskip}{0.0cm}
i_u^* = \arg \max _{i \in \mathcal{I}} P\left(i_{m+1}^{(u)}=i \mid s_u\right).
\end{equation}

\subsection{Attacker Brief}
\label{sec:Threat Model}
Building on the base SRS described above, this section examines the attacker from three different perspectives.

\noindent\textbf{Attacker's Goal.}
Poisoning attacks in RSs are categorized into non-targeted and targeted attacks \cite{fan2022survey, tian2022comprehensive}. Non-targeted attacks focus on degrading the overall system performance \cite{wu2023influence,wu2022fedattack}, whereas targeted attacks aim to promote or demote specific items \cite{wang2024llm,wu2022poisoning}. This work focuses on targeted promotion attacks, aiming to boost the visibility of target items across users' recommendation list.

\noindent\textbf{Attacker's Background Knowledge.}
Due to security and privacy reasons, it is impractical to assume any access to any SRS's architecture. As such, this work focuses on black-box attacks with no direct access to the victim model. As for data, in practical SRS settings, a visitor (either a benign user or attacker) commonly has partial access to the interaction records of other users. For example, on e-commerce sites like Amazon and eBay, the accessible interactions include product reviews/ratings voluntarily published by a fraction of the users $\mathcal{U}' \subset \mathcal{U}$. Each user $u'\in \mathcal{U}'$ publishes a genuine interaction sequence denoted as $s_{u'}$, and $\mathcal{S}'=\{s_{u'}|u'\in \mathcal{U}'\}\subset \mathcal{S}$ is the collection of all public sequences. Meanwhile, the majority part of the interactions $\mathcal{S}$ remains unavailable to the attacker. 

\noindent\textbf{Attacker's Capability.}
Attackers can inject malicious users $\tilde{u}$ with forged interactions into the SRS, which become a part of its training data and mislead the victim model. Due to resource limitations, we assume that the attacker can register only a small set of fake users $\mathcal{U}_F$, e.g., $\frac{|\mathcal{U}_F|}{|\mathcal{U}|}=$ 1\% -- a common setup in related work \cite{wang2023poisoning, wang2023revisiting}. Additionally, attackers can query the black-box victim model to obtain corresponding recommendation outputs, allowing them to refine and adjust the attack mechanism.

\subsection{Poisoning Attack: Problem Definition}
\label{An Optimization Problem}
To attack the victim recommendation model, each fake user $\tilde{u} \in \mathcal{U}_F$ generates one interaction sequence $s_{\tilde{u}} = \{ i_m^{(\tilde{u})}\}_{m=1}^M$ ($i_m^{(\tilde{u})} \in \mathcal{I}$), termed poisoning sequence. The full set of poisoning sequences \(\mathcal{S}_F\) joins all benign users' interactions $\mathcal{S}$. The victim model, denoted by $f_{\theta}(\cdot)$ is then updated with $\mathcal{S}_F \cup \mathcal{S}$. Ideally, with a carefully crafted $\mathcal{S}_F$, the updated $f_{\theta}(\cdot)$ can exhibit the attacker's designated behaviors, e.g., recommending a specific item more frequently.

Intuitively, to ensure the poisoning sequences are effective, the attacker needs to iteratively optimize every $s_{\tilde{u}}\in \mathcal{S}_F$ based on how $f_{\theta}(\cdot)$ responds to the previously generated $\mathcal{S}_F$. However, it is impractical to intensively query the victim model at scale, and the black-box attack setting further prevents the attacker from obtaining a local copy of it. Hence, as a common practice, a surrogate, white-box model $f_w(\cdot)$ is built \cite{fan2021attacking,nguyen2023poisoning} by the attacker to simulate the victim model. Following Section \ref{sec:Threat Model}, let \(\mathcal{S'}\subset \mathcal{S}\) denote the publicly accessible user-item interaction sequences, based on which the surrogate model is trained. Note that we do not particularly discuss other alternatives for obtaining the surrogate model, since our main innovation lies in subsequent attack steps and there exists a separate line of work on model extraction attacks \cite{zhang2021reverse,yue2021black,zhu2023membership}. By training $f_w(\cdot)$ on \(\mathcal{S'}\), it is able to approximate the behavioral patterns of the victim model trained on \(\mathcal{S}\) 
\cite{tang2020revisiting}.
By substituting $f_{\theta}(\cdot)$ with $f_w(\cdot)$, the poisoning attack can be formulated as a bi-level optimization problem:
\vspace{-1mm}
{
\begin{align}
 & \arg \max _{\mathcal{S}_F} \mathcal{L}_{atk}(f_{w^*}(\mathcal{S}'\cup \mathcal{S}_F)),\nonumber \\ & \text{s.t.} \: w^* = \arg \min _{w} \mathcal{L}_{rec}\left(f_{w}\left(\mathcal{S}'\cup \mathcal{S}_F\right)\right).
\label{equ:outer}
\end{align}}

Here, \(\mathcal{L}_{rec}\) is the recommendation objective for the \textit{inner optimization}, which obtains the optimal model parameter \(w^*\) by minimizing the prediction error based on interaction sequences $\mathcal{S}'\cup \mathcal{S}_F$. Meanwhile, $\mathcal{L}_{atk}$ is the attack objective for the \textit{outer optimization}, which steers the update of $\mathcal{S}_F$. Specifically, the attacker aims to push a target item $t \in \mathcal{I}$ into as many users' top-\(K\) lists as possible, and the design choices of $\mathcal{L}_{atk}$ will be unfoleded in Section \ref{part1}.

\section{Methodology}
\subsection{Overview of DDSP}
\begin{figure*}
    \setlength{\abovecaptionskip}{0.0cm}
    \setlength{\belowcaptionskip}{0.0cm}
    \centering
    \includegraphics[width=\linewidth]{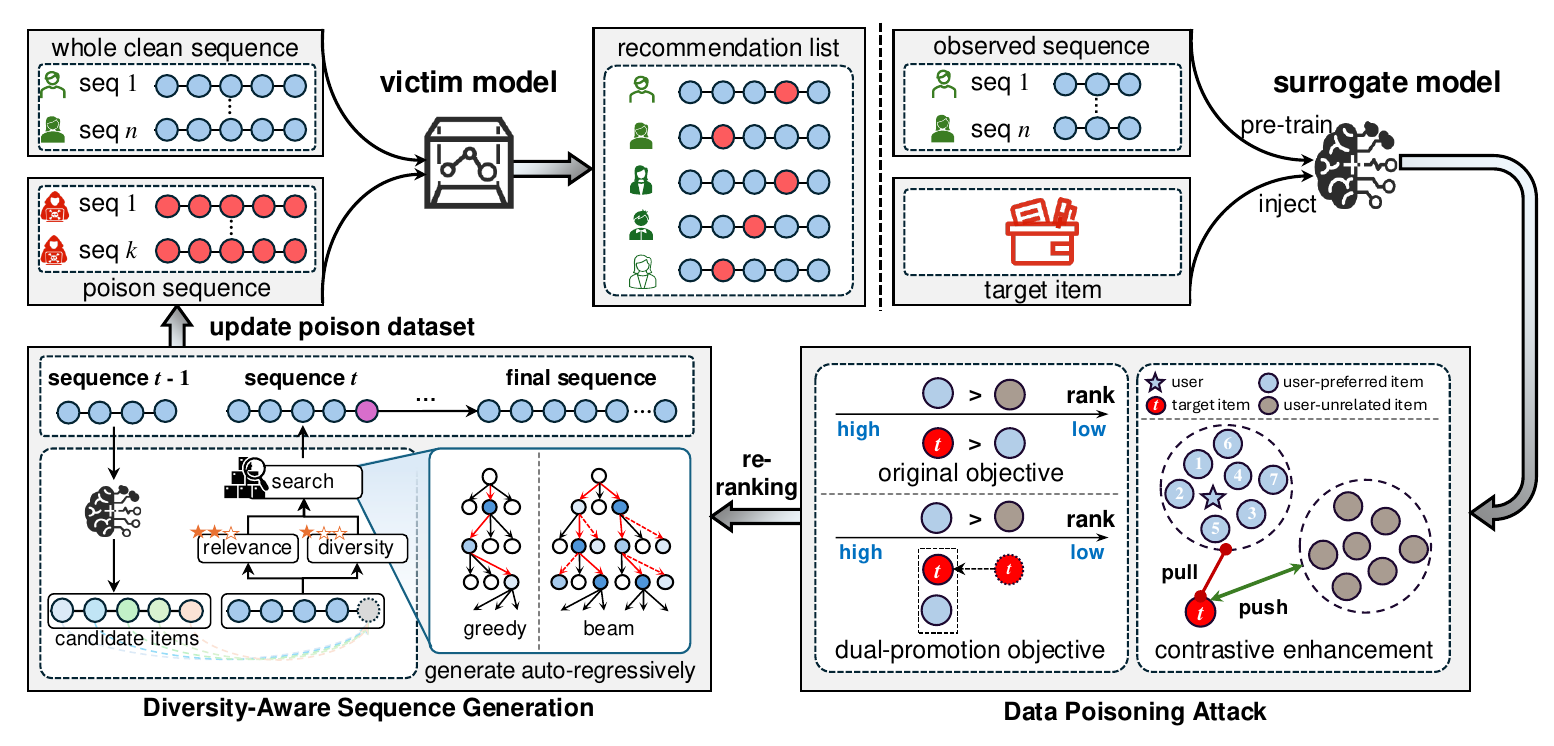}
    \caption{The framework of our proposed DDSP.}
    \label{framework}
\end{figure*}

Our framework, namely Diversity-aware Dual-promotion Sequential Poisoning attack (DDSP), is depicted in Fig. \ref{framework}. DDSP has two novel designs: (1) the dual-promotion attack objective (Section \ref{part1}) that boosts the target item's exposure with minimal harms to the recommendation accuracy; and (2) the diversity-aware sequence generation (Section \ref{part3}) that improves the diversity of the poisoning sequences. Both components are used along with the surrogate model $f_w(\cdot)$. The following sections detail the design of DDSP.

\subsection{\textbf{Dual-Promotion Attack Objective}}
\label{part1}

As described in Section \ref{An Optimization Problem}, the bi-level optimization framework involves two objectives $\mathcal{L}_{rec}$ and $\mathcal{L}_{atk}$. For $\mathcal{L}_{rec}$, one of the most popular objective is Bayesian Personalized Ranking (\(\textit{BPR}\)) loss \cite{rendle2012bpr}, which is defined as follows:
\begin{equation}
\label{BPR}
\mathcal{L} _{rec-BPR}=-\sum_{(u, i^+, i^-) \in \mathcal{D}} \log \sigma\left(\hat{y}_{u i^+}-\hat{y}_{u i^-}\right),
\end{equation}
where \(\sigma(\cdot)\) is the logistic sigmoid function, \(\mathcal{D}\) is the set of triplets (\(u,i^+,i^-\)), with \(i^+\) being a positive (i.e., visited) item and \(i^-\) a negative (i.e., unvisited) item for $u$. \(\hat{y}_{u i}=\mathbf{e}_u^{\top} \mathbf{e}_{i}\) is the model's prediction on $u$'s preference score for an arbitrary item \(i\). Note that in SRSs, a user's preference embedding $\mathbf{e}_u$ is commonly generated by a sequence encoder $f_{enc}(\cdot)$ that fuses all item embeddings $\mathbf{e}_i$ in the interaction sequence \cite{kang2018self,zhou2020s3}, i.e., $\mathbf{e}_u = f_{enc}(\{\mathbf{e}_i|i\in s_u\})$. Also, as $\mathcal{L}_{rec}$ is used for training the surrogate model $f_w(\cdot)$, $\mathcal{D}$ is constructed based on $\mathcal{S}'\cup \mathcal{S}_F$.

The core idea of a recommendation loss is to align the embeddings of users and their interacted positive items. Taking BPR loss as an example, the gradient of the positive item embedding \(\mathbf{e}_{i^+}\) is:
\begin{equation}\label{eq:BPRgrad}
    \frac{\partial \mathcal{L}_{{rec-BPR}}}{\partial \mathbf{e}_{i^+}}=-\!\!\!\!\! \sum_{(u, i^+, i^-) \in \mathcal{D}}\!\!\!\!\! \left(1-\sigma\left(\Delta\right)\right) \mathbf{e}_u, \,\, \Delta=\mathbf{e}_u^{\top} \mathbf{e}_{i^+} - \mathbf{e}_u^{\top} \mathbf{e}_{i^-},
\end{equation}
which translates into a gradient update step for $\mathbf{e}_i$ (with learning rate $\alpha$) of: \(\mathbf{e}_{i^+} \leftarrow \mathbf{e}_{i^+} - \alpha \cdot \frac{\partial \mathcal{L}_{\mathrm{BPR}}}{\partial \mathbf{e}_{i^+}} = \mathbf{e}_{i^+} + \alpha \cdot \sum_{(u, i^+, i^-) \in \mathcal{D}}\left(1-\sigma\left(\Delta\right)\right) \mathbf{e}_u\). Intuitively, for positive item $i^+$, \(\mathbf{e}_{i^+}\) gradually moves closer to \(\mathbf{e}_u\) during training. In contrast, for the negative item $i^-$, its embedding update trajectory is \(\mathbf{e}_{i^-} \leftarrow \mathbf{e}_{i^-} - \alpha \cdot \frac{\partial \mathcal{L}_{\mathrm{BPR}}}{\partial \mathbf{e}_{i^-}} = \mathbf{e}_{i^-} - \alpha \cdot \sum_{(u, i^+, i^-) \in \mathcal{D}}\left(1-\sigma\left(\Delta\right)\right) \mathbf{e}_u\), forcing it to move in the opposite direction from \(\mathbf{e}_u\). Another popular choice for $\mathcal{L}_{rec}$ is the binary cross-entropy (BCE) loss. By dedicating notations $\hat{y}_{u i^+}$ and $\hat{y}_{u i^-}$ to predicted scores respectively on positive and negative items, it can be written as the following:
\begin{equation}
\setlength{\abovecaptionskip}{0.0cm}
\setlength{\belowcaptionskip}{0.0cm}
\begin{aligned}
\label{BCE}
\mathcal{L} _{rec-BCE} = - \sum_{(u, i^+, i^-) \in \mathcal{D}}(\log(\hat{y}_{ui^+}) + \log (1 - \hat{y}_{ui^-})).
\end{aligned}
\end{equation}
With a gradient-based analysis that is analogous to Eq. (\ref{eq:BPRgrad}), the same conclusion on embeddings' update process can be quickly drawn.

As for $\mathcal{L}_{atk}$, the goal of item promotion is to increase the appearance of a target item \(t \in \mathcal{I}\) in generated recommendation results. A common practice is to boost $t$'s exposure in users' top-$K$ recommendation lists, measured by Hit Ratio (HR@\(K\)) \cite{lin2022shilling,li2016data,fang2020influence}:
\begin{equation}\label{eq:HRK_t}
\text{max} \quad HR@K = \frac{1}{\left|\mathcal{U}\right|} \sum_{u \in \mathcal{U}} \mathbb{I}\left(t, \Gamma_u\right) \text{, }
\end{equation}
where \(\mathbb{I}(\cdot)\) is an indicator function that returns 1 if item \(t\) appears in user \(u\)'s top-\(K\) recommendation list \(\Gamma_u\), and 0 otherwise. To achieve Eq. (\ref{eq:HRK_t}), the attacker searches for an optimal poisoning sequence $s_{\tilde{u}}$ for each fake user \(\tilde{u}\), such that the recommendations are altered after model update. Given the NP-hard nature of optimizing all $s_{\tilde{u}}$ and the unavailability of all benign users' recommendation lists, directly optimizing is infeasible. Therefore, a workaround \cite{huang2021data,wu2024accelerating} is to instead maximize the score $\hat{y}_{ut}$ between $u$ and $t$ generated by the surrogate model, such that the surrogate model is more likely to rank $t$ at a higher place. Bearing this intuition, a common attack objective \cite{tang2020revisiting,wang2023revisiting} is to promote \(t\)'s rank among the full item list $\mathcal{I}$: 
\begin{equation}
\mathcal{L}_{atk-list} = -\sum_{u \in \mathcal{U}'\cup \mathcal{U}_F} \log \left( \frac{\exp(\hat{y}_{ut})}{\sum_{i \in \mathcal{I}} \exp(\hat{y}_{ui})} \right).
\label{atk_obj1}
\end{equation}
To avoid the efficiency hurdle from full-list item ranking, a widely used alternative \cite{huang2021data} ensures \(t\) surpasses the items in each users' top-\(K\) list \(\Gamma_u\), which is formulated in a pairwise fashion:  
\begin{equation}
\mathcal{L}_{atk-pair} = \sum_{u \in \mathcal{U}'\cup \mathcal{U}_F} \sum_{i \in \Gamma_u} g(\hat{y}_{ui} - \hat{y}_{ut}),
\label{atk_obj2}
\end{equation}
where $g(x)=\frac{1}{1+\exp (-x / b)}$ is the Wilcoxon-Mann-Whitney function~\cite{backstrom2011supervised} with a width parameter $b$ that ensures that $\mathcal{L}_{atk-pair} \geq 0$ and is differentiable. 
  
\begin{figure}
    \setlength{\abovecaptionskip}{0.0cm}
    \setlength{\belowcaptionskip}{0.0cm}
    \centering
    \includegraphics[width=\linewidth]{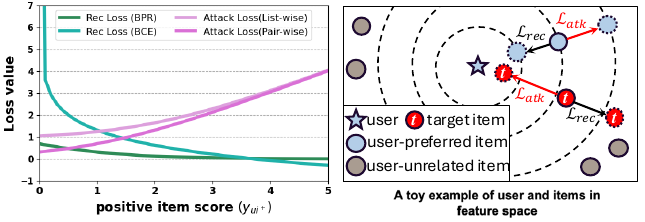}
    \caption{(Left) Comparison of the training curves for two recommendation objectives -- \(\mathcal{L}_{rec-BPR}\) and \(\mathcal{L}_{rec-BCE}\) (green lines) and two attack objectives -- \(\mathcal{L}_{atk-list}\) and \(\mathcal{L}_{atk-pair}\) (purple lines), as the positive item score \(y_{ui^+}\) increases. We assume non-negativity of all embeddings. For visualization purpose, we set one each for positive,  negative, and target items, and we fix the scores of target and negative items respectively to $\hat{y}_{ut}=1$ and $\hat{y}_{ui^-}=0$. (Right) A toy example showing how the user, target item, and user‐preferred items move in the feature space under the influence of the recommendation objective and the attack objective.}
    \label{loss_compare}
\end{figure}

\textbf{The Conflicting Goals of $\mathcal{L}_{atk}$ and $\mathcal{L}_{rec}$.} For both attack losses, the higher \(\hat{y}_{ut}\), the smaller the loss value. However, represented by $\mathcal{L}_{atk-list}$ and $\mathcal{L}_{atk-pair}$, existing attack objectives can degrade the recommendation performance as they inadvertently pushes each user's preferred items (especially the interacted positive items $i$) away from the user \(u\) in the embedding space. Taking \(\mathcal{L}_{atk-pair}\) as an example, we first derive the gradient of the target item embedding $\mathbf{e}_t$: 
\begin{align}
    &\frac{\partial \mathcal{L}_{atk-pair}}{\partial \mathbf{e}_t}= \sum_{u\in \mathcal{U}'\cup \mathcal{U}_F} \sum_{i \in \Gamma_u} \frac{\partial g\left(\Delta \right)}{\partial \mathbf{e}_t} = -\sum_{u\in \mathcal{U}'\cup \mathcal{U}_F} \sum_{i \in \Gamma_u}g^{\prime}\left(\Delta\right) \mathbf{e}_u,\nonumber\\
    &\Delta=\hat{y}_{ui} - \hat{y}_{ut}, \,\,\,\, g^{\prime}(\Delta) = \frac{1}{b} g(\Delta)[1-g(\Delta)],
\end{align}
hence the update rule for \(\mathbf{e}_t\) with learning rate $\alpha$ is: 
\begin{equation}
    \mathbf{e}_t \leftarrow \mathbf{e}_t-\alpha \cdot \frac{\partial \mathcal{L}_{{atk-pair}}}{\partial \mathbf{e}_t} = \mathbf{e}_t+\alpha \cdot \sum_{u\in \mathcal{U}'\cup \mathcal{U}_F} \sum_{i \in \Gamma_u} g^{\prime}\left(\Delta\right) \mathbf{e}_u .
\end{equation}
 Because \(g^{\prime}(\Delta) > 0\), \(\mathbf{e}_t\) moves towards \(\mathbf{e}_u\) during training, contributing to the increase of $t$'s ranking score $\hat{y}_{ut}$. Unfortunately, the same conclusion cannot be made for the embedding of positive item $\mathbf{e}_i$. For any item \(i \in \Gamma_u\), the gradient is \(\frac{\partial \mathcal{L}_{{atk-pair}}}{\partial \mathbf{e}_i}= \sum_{u\in \mathcal{U}'\cup \mathcal{U}_F} \sum_{i \in \Gamma_u} g^{\prime}\left(\Delta\right) \mathbf{e}_u\), leading to the following update for \(\mathbf{e}_i\): 
 \begin{equation}
     \mathbf{e}_i \leftarrow \mathbf{e}_i- \frac{\partial \mathcal{L}_{{atk-pair}}}{\partial \mathbf{e}_i} =  \mathbf{e}_i-\alpha \cdot \sum_{u\in \mathcal{U}'\cup \mathcal{U}_F} \sum_{i \in \Gamma_u} g^{\prime}\left(\Delta\right) \mathbf{e}_u.
 \end{equation}
  Essentially, \(\mathbf{e}_i\) is pushed away from \(\mathbf{e}_u\) for $i\in\Gamma_u$ and $i\neq t$, including any positive item(s) $i^+$ that are highly likely to appear in $\Gamma_u$ after the inner optimization towards $\mathcal{L}_{rec}$. As a result, it conflicts with the recommendation objective. To better illustrate such a conflict, on the left of Fig. \ref{loss_compare}, we plot the loss curves of \(\mathcal{L}_{rec-BCE}\), \(\mathcal{L}_{rec-BPR}\), \(\mathcal{L}_{atk-list}\), and \(\mathcal{L}_{atk-pair}\) w.r.t. the model-generated score \(\hat{y}_{ui^+}\) of a positive user-item pair $(u, i^+)$. The right panel of Fig. \ref{loss_compare} offers a more intuitive depiction of how items move in the embedding space. When learning the embeddings of a user's interacted item $\mathbf{e}_{i^+}$, the contradicting goals of $\mathcal{L}_{atk}$ and $\mathcal{L}_{rec}$ tend to make $\hat{y}_{ui^+}$ stall on a point that is suboptimal for either objective. This also explains the need for the bi-level optimization in Eq.(\ref{equ:outer}) instead of the more efficient joint optimization, as the direct combination of $\mathcal{L}_{rec}$ and $\mathcal{L}_{atk}$ will heavily hamper the informativeness of $\mathbf{e}_{i^+}$'s gradient during optimization.

\textbf{Our Solution.} Rather than inflating the score of the target item above all others, we aim to bring the target item's embedding representation closer to those items the user genuinely prefers (e.g., items in \(\Gamma_u\)) such that both sides are simultaneously promoted, fulfilling the attacker's objective while preserving recommendation accuracy and stealthiness. Thus, we propose to minimize the distance between \(t\) and all \(i\in \Gamma_u\) via the following:
\begin{equation}
\mathcal{L}_{atk}= \sum_{u\in \mathcal{U}'\cup \mathcal{U}_F}\sum_{i \in \Gamma_u}\left(\hat{y}_{ui}-\hat{y}_{ut}\right)^2.
\label{l_atk}
\end{equation}

The Mean Squared Error (MSE) objective can promote both the target item \(t\) and the user preferred items \(i \in \Gamma_u\) simultaneously.  Specifically, let \(\Delta= \hat{y}_{ui}-\hat{y}_{ut} = \mathbf{e}_u^{\top} \mathbf{e}_i - \mathbf{e}_u^{\top} \mathbf{e}_t\), then \(\frac{\partial \mathcal{L}_{{atk}}}{\partial \mathbf{e}_i} = 2 \sum_{u\in \mathcal{U}'\cup \mathcal{U}_F}\sum_{i \in \Gamma_u} \Delta \mathbf{e}_u\). Under gradient descent with learning rate \(\alpha\), \(\mathbf{e}_i \leftarrow \mathbf{e}_i-\alpha \frac{\partial \mathcal{L}_{atk}}{\partial \mathbf{e}_i}=\mathbf{e}_i-2\alpha \sum_{u\in \mathcal{U}'\cup \mathcal{U}_F}\sum_{i \in \Gamma_u} \Delta \mathbf{e}_u \). When \(\Delta < 0\) (i.e., \(\hat{y}_{ui} < \hat{y}_{ut}\)), \(\mathbf{e}_i\) moves in the same direction as \(\mathbf{e}_u\), reducing the difference between item \(i\) and user \(u\), then boosting \(\hat{y}_{ui}\). Otherwise, \(\hat{y}_{ut}\) increases. Similarly, we can derive \(\mathbf{e}_t \leftarrow \mathbf{e}_t+2\alpha\sum_{u\in \mathcal{U}'\cup \mathcal{U}_F}\sum_{i \in \Gamma_u} \Delta \mathbf{e}_u\). When \(\Delta > 0\) (i.e., \(\hat{y}_{ui} > \hat{y}_{ut}\)), \(\mathbf{e}_t\) moves towards \(\mathbf{e}_u\), reducing the difference between target item \(t\) and user \(u\), thus increasing \(\hat{y}_{ut}\). Otherwise, \(\hat{y}_{ui}\) increases. This dual adjustment aligns both \(\mathbf{e}_i\) and \(\mathbf{e}_t\) with \(\mathbf{e}_u\), enabling the target item \(t\) to be promoted without harming genuine preferences. Furthermore, as we will discuss in Section \ref{sec:MT}, this revamped $\mathcal{L}_{atk}$ can help DDSP bypass the complex bi-level optimization in Eq.(\ref{equ:outer}) and instead jointly optimize $\mathcal{L}_{atk}$ and $\mathcal{L}_{rec}$. 

\textbf{Contrastive Regularization.} To further increase the likelihood that both the target item \(t\) and user preferred items \(i \in \Gamma_u\) are recommended, a contrastive regularization term extends beyond MSE, which only shrinks the gap between \(t\) and \(i \in \Gamma_u\) without addressing irrelevant items. Contrastive learning \cite{wang2025federated} instead aligns \(t\) with items in \(\Gamma _u\) while pushing it away from negatives. Treating all \(i \in 
\Gamma_u\) as $t$'s positive counterparts and all other items as negative ones, the contrastive regularizer is defined as follows:
\begin{equation}
\mathcal{L}_{reg}=-\sum_{i \in \Gamma_u} \log \frac{\exp \left({\mathbf{e}}_t{ }^{\top} {\mathbf{e}}_i / \tau\right)}{\sum_{k \in \mathcal{I}} \exp \left({\mathbf{e}}_t^{\top} {\mathbf{e}}_k / \tau\right)},
\label{l_nce}
\end{equation}
where \(\mathbf{e}_t\), \(\mathbf{e}_i\), and \(\mathbf{e}_k\) are the \(L2\)--regularized embeddings of the target item \(t\), user preferred items \(i\) in \(\Gamma _u\), and other items, respectively. This regularizer pulls the target item to user preferred items and pushes it away from less relevant ones, ensuring it appears more frequently alongside user preferred items and preventing irrelevant items from overshadowing it.

\vspace{-3mm}
\subsection{\textbf{Diversity-Aware Sequence Generation}}
\label{part3}
Dual-promotion attack objective is designed to guide the generation of poisoning sequences using $f_w(\cdot)$. To make sure the generated sequences encode temporal dependencies, we adopt an auto-regressive generation strategy \cite{yue2021black}, thereby simulating user behaviors realistically.

Fig. \ref{framework} (lower left) illustrates the process of generating poisoning sequences for fake users. Specifically, to generate a sequence \(s_{\tilde{u}}\), we randomly sample an initial item \(\{i_{1}^{(\tilde{u})}\}\) and feed it to the well-trained \({f}_w(\cdot)\), which outputs a recommendation list \(\Gamma _{\tilde{u}} = \boldsymbol{f}_w (i_{1}^{(\tilde{u})})\). A sampling mechanism then picks the subsequent item from \(\Gamma _{\tilde{u}}\), which extends $s_{\tilde{u}}$ by one. This process repeats until \(s_{\tilde{u}}\) reaches the maximum length \(M\), formally defined as:
\begin{equation}
i_{m}^{(\tilde{u})} = \text{sampler} \left( {f}_w(\{i_{1}^{(\tilde{u})},i_{2}^{(\tilde{u})},..., i_{m-1}^{(\tilde{u})}\})\right), \,\,\,\, m = 1, 2, ..., M,
\end{equation}
where \text{sampler(\(\cdot\))} samples one item from the given top-\(K\) list. As such, we generate \(|\mathcal{U}_F|\) fake sequences \(\mathcal{S}_F\). 

Although the described method generates in-distribution sequences \cite{jiang2024memory}, it has a key shortcoming: as the surrogate model is trained, the target item appears more frequently in $s_{\tilde{u}}$ due to the strong preference signal for the target item (and any user preferred items), with no mechanisms to maintain diversity in the generated poisoning sequences. As a result, \(\Gamma_{\tilde{u}}\) (hence $s_{\tilde{u}}$) is prone to collapsing towards a naive solution where the target item appears excessively. This in turn harms the stealthiness, and will eventually impede the success of the attack. 

To address this, a re-ranking-based strategy \cite{kelly2006enhancing,zhao2023fairness} is introduced to enhance sequence diversity. In our case, when generating the next item in $s_{\tilde{u}}$, re-ranking methods adjust the top-pick from the ranking list $\Gamma_{\tilde{u}}$ produced by the surrogate model, by jointly considering the diversity and utility (i.e., relevance) of the expanded $s_{\tilde{u}}$. To facilitate this, a sequence-level score is defined:
\begin{equation}
r\left(s_{\tilde{u}}\right)=\frac{1-\lambda}{\left|s_{\tilde{u}}\right|} \sum_{i \in s_{\tilde{u}}} \hat{y}_{ui}+\lambda {f}_{{div}}\left(s_{\tilde{u}}\right),
\label{relevance-diversity}
\end{equation}
where \({f}_{{div}}(s_{\tilde{u}})\) indicates the diversity score of the current poisoning sequence \(s_{\tilde{u}}\), and \(\lambda\) controls the trade-off between utility and diversity. The diversity metric is defined as follows:
\begin{equation}{f}_{{div}}\left(s_{\tilde{u}}\right)=\frac{1}{\left|s_{\tilde{u}}\right|\left(\left|s_{\tilde{u}}\right|-1\right)} \sum_{i \in s_{\tilde{u}}} \sum_{j \in s_{\tilde{u}} \backslash i} d(i, j),
\end{equation}
where \(d(i,j)\) measures the dissimilarity or distance between items \(i\) and \(j\). In this paper, \(L2\)-norm \cite{cortes2012l2} is used to calculate the distance between learned item embeddings. 

\begin{algorithm}[t]
\caption{Diversity-Aware Sequence Generation}
\label{alg:diversity}
\begin{algorithmic}[1]
\State \textbf{Input:} Surrogate model $f_w(\cdot)$; Beam width \(B\); Diversity weight $\lambda$; Maximum length of sequence \(M\).
\State \textbf{Output:} Generated fake sequences \(\mathcal{S}_F\).
\State Set \(\mathcal{S}_F \leftarrow \emptyset \), randomly initialize root node \(Beam \leftarrow [rand] \);  
\ForEach{\(\tilde{u} \in \mathcal{U}_F\)}
    \State $s \leftarrow \{i_{rand}\}$; \Comment{Random initialization with $i_{rand}\in\mathcal{I}$}
    \State \(Beam \leftarrow \{(s, r(s))\}\); \Comment{Initializing Beam}
    \While{\(|s| < M\)}
        \State \(TempBeam \leftarrow \emptyset\);
        \ForEach{\((s, r(s)) \in Beam\)}
            \State $\Gamma_{\tilde{u}} \leftarrow f_w(s)$;
            \ForEach{\(i \in {\Gamma_{\tilde{u}}} \setminus s\)}
                \State \(s_{new} \leftarrow s \cup i\);
                \State \(TempBeam \leftarrow TempBeam \cup (s_{new}, r(s_{new})) \);
            \EndFor
        \EndFor
        \State \(Beam \gets \)  Top-\(B(TempBeam)\);
        \Comment{Keep top \(B\) sequences}
    \EndWhile    
    \State \(s_{\tilde{u}} = \arg\max_{s \in Beam} r(s)\); \Comment{Make the best sequence $s_{\tilde{u}}$}
    \State \(\mathcal{S}_F =  \mathcal{S}_F \cup s_{\tilde{u}}\);
\EndFor
\end{algorithmic}
\end{algorithm}

To expand $s_{\tilde{u}}$ under the guidance of $r(s_{\tilde{u}})$, two search methods are employed to generate sequences. One is a straightforward greedy Maximum Marginal Relevance (MMR) algorithm \cite{carbonell1998use}. It is performed iteratively to select the item that can maximize the current $r\left(s_{\tilde{u}}\right)$ until the expected sequence length is reached. This greedy algorithm trades off some performance and gains strong efficiency in return. To pursue better optimality of the final poisoning sequences, we further enhance the greedy algorithm with beam search, whose process is shown in Algorithm \ref{alg:diversity}. In short, it selects a set of items with the highest scores as candidates for the next step. These candidates are expanded iteratively, allowing the search to consider multiple potential paths in a tree. The process continues until the sequence reaches the expected length. Notably, when the beam width is \(B=1\), it reverts back to the basic greedy algorithm. 

\subsection{Model Training}\label{sec:MT}
A summary of DDSP's attack pipeline is shown in Algorithm \ref{alg:DDSP}. After initialization, once the surrogate model $f_w(\cdot)$ has generated  poisoning sequences \(\mathcal{S}_F\) for all fake users (line 6), \(\mathcal{S}_F\) will be used in combination with $\mathcal{S}'$ to further update $f_w(\cdot)$ towards both recommendation and attack objectives (line 7). This iterative process will continue until the overall loss $\mathcal{L}$ converges or is sufficiently small. Because the attack objective no longer conflicts with the recommendation objective, DDSP brings the extra convenience of replacing bi-level optimization with joint training. Hence, we fuse recommendation loss \(\mathcal{L}_{rec}\), dual-promotion attack loss \(\mathcal{L}_{atk}\), and contrastive loss \(\mathcal{L}_{reg}\) in the overall optimization objective $\mathcal{L}$:
\vspace{-1mm}
\begin{equation}
\setlength{\abovecaptionskip}{0.0cm}
\setlength{\belowcaptionskip}{0.0cm}
\mathcal{L} = \mathcal{L}_{rec} + \mathcal{L}_{atk} + 
\eta \mathcal{L}_{reg},
\label{l_all}
\end{equation} 
where we adopt the binary cross-entropy (Eq.(\ref{BCE})) as the recommendation objective $\mathcal{L}_{rec}$, and $\eta$ controls the effect of the contrastive regularization \cite{jiang2024challenging}. Once the poisoning sequences are fully optimized, they are then injected into the black-box victim model to promote the target item $t$.

Throughout the entire process of generating quality poisoning sequences, the attacker has only had access to a small portion of real user interactions $\mathcal{S}'$. In DDSP, the revamped attack objective is able to mitigate negative impact on the recommendation task, and the generated poisoning sequences are further enhanced with diversity-awareness. These two components collaboratively contribute to effective targeted item promotion, while lowering DDSP's detectability during attacks with reduced traces. Furthermore, the optimization of $\mathcal{S}_F$ do not require any intermediate feedback from the black-box victim model by querying it, which avoids the potential communication overhead.

\begin{algorithm}[t]
\caption{Attack Pipeline of DDSP}
\label{alg:DDSP}
\begin{algorithmic}[1]
\State \textbf{Input:} Observed sequence \(\mathcal{S'}\); Surrogate model \({f}_w(\cdot)\); Fake user set \(\mathcal{U}_F\); Target item \(t\).
\State \textbf{Output:} Poisoning sequences \(\mathcal{S}_F\).
\State Pre-train \(f_w(\cdot)\) with $\mathcal{S'}$ only w.r.t. $\mathcal{L}_{rec}$;
\State \(\mathcal{S}_F \leftarrow \emptyset\);
\While{not converged}
\State \(\mathcal{S}_F \leftarrow\) Algorithm \(\ref{alg:diversity}\);
\State Re-train \({f}_w(\cdot)\) with (\(\mathcal{S'} \cup \mathcal{S}_F\) ) w.r.t. $\mathcal{L}$ (Eq. (\ref{l_all}));
\EndWhile \Comment{$\mathcal{S}_F$ will be used for poisoning the victim model}
\end{algorithmic}
\end{algorithm}

\section{Experiments}
In this section, we conduct extensive experiments on three datasets to validate the effectiveness of DDSP.

\subsection{Setup}
\subsubsection{\textbf{Datasets}}
We evaluate the proposed attack method on three real-world Amazon datasets \cite{zhou2020s3,wang2023poisoning}: \textit{Beauty}, \textit{Sports and Outdoors} (short for Sports), and \textit{Toys and Games} (short for Toys), each representing public customer interactions. Following the data pre-processing steps in \cite{wang2023poisoning}, we exclude all users and items with fewer than five interactions. Table \ref{dataset statistics} summarizes the processed datasets.
\vspace{-2mm}
\begin{table}[H]
    \setlength{\abovecaptionskip}{0.0cm}
    \setlength{\belowcaptionskip}{0.0cm}
    \setlength{\tabcolsep}{2pt}
  \caption{Dataset statistics}
  \label{dataset statistics}
  \begin{tabular}{c|ccc}
    \toprule
    \textbf{Datasets} & \textbf{Beauty} & \textbf{Sports and Outdoors} & \textbf{Toys and Games}\\
    \midrule
    Users & 22,363 & 35,598 & 19,412 \\
    Items & 12,101 & 18,357 & 11,924\\
    Interactions & 198,502 & 296,337 & 167,597\\
  \bottomrule
\end{tabular}
\end{table}
\vspace{-4mm}
\subsubsection{\textbf{Base Recommender and Baselines}}
To validate the effectiveness of DDSP in SRSs, we use SASRec \cite{kang2018self} as the surrogate model. To demonstrate the universality of DDSP across different black-box victim models, we evaluate two victim models: SASRec, which represents the scenario where the surrogate and victim models share highly similar or even identical structures, and CL4SRec \cite{xie2022contrastive}, a popular SRS that differs from the surrogate model, allowing us to test the adaptability of our method to structurally distinct models. Two types of attack models are selected as our baselines: \textit{(1) Heuristic-based methods.} \textbf{Pure} shows the normal performance without attack. \textbf{Random Attack} \cite{lin2022shilling} proposes to interact with the target item and other available items randomly. \textbf{Bandwagon Attack} \cite{li2016data} injects fake co-visitations between popular items and the target item. \textit{(2) Optimization-based methods.} \textbf{GTA} \cite{wang2023revisiting} encourages fake users to interact with items predicted to have high ratings and the target item. \textbf{CLeaR} \cite{wang2024unveiling} introduces a smoother spectral value distribution to amplify the contrastive loss's inherent dispersion effect. \textbf{SSL} \cite{wang2023poisoning} is a poisoning attack method against self-supervised learning-based SRS. It utilizes a generative adversarial network to generate fake sequences.

\subsubsection{\textbf{Implementation Details.}}
We adopt the public PyTorch implementations of SASRec and CL4SRec from \cite{yu2023self}, optimizing all models with Adam optimizer \cite{kingma2014adam}. The learning rate is set to 0.001, batch size to 256, dropout rate to 0.2, and item embedding size to 64. The maximum sequence length \(M\) is set to 50. 
For attack models, we use the public implementation of {GTA} and {CLeaR} from \cite{wang2024unveiling}. The malicious user budget is 1\% of the normal user base (i.e., $\frac{|\mathcal{U}_F|}{|\mathcal{U}|}=1\%$), with each malicious user having the same average number of interactions as a normal user. We choose one unpopular item from the least-popular set as the target item. 
The surrogate model shares the same architecture as SASRec, pre-trained with 10\% of the real interactions to simulate practical scenarios. We set the weight $\eta$ of contrastive regularizer $\mathcal{L}_{reg}$ to $0.01$ after searching within \{0.01, 0.05, 0.1, 0.15, 0.2\}, its temperature coefficient \(\tau=0.2\) after searching within \{0.1, 0.2, 0.3, 0.4, 0.5\}. We evaluate performance with two commonly used metrics in RSs: Hit Ratio \cite{li2022revisiting} and NDCG \cite{zhang2023revisiting}. When evaluating recommendation performance, the last item in each sequence is treated as the relevant (ground truth) item. For attack performance, the target item is deemed relevant, and we measure both the frequency and the ranking of the target item in real users' recommendation lists. Higher scores indicate more successful promotion of the target item. Each experiment is conducted 5 times, and the average results are reported.

\begin{table*}[]
\setlength{\abovecaptionskip}{0.0cm}
\setlength{\belowcaptionskip}{0.0cm}
\setlength{\tabcolsep}{3pt}
\small
\caption{We evaluate the impact of baselines and DDSP on two backbone models: SASRec \cite{kang2018self} and CL4SRec \cite{xie2022contrastive}. The attack budget is 1\% in this table. \textit{Pure} indicates performance with no attacks. \textit{Rec} is recommendation performance, and \textit{Atk} is attack performance (shaded columns). Metrics \textit{H} and \textit{N} denote Hit Ratio and NDCG, respectively. DDSP-G and DDSP-B are variants using greedy and beam search, respectively. The best performance is bolded, and the second-best is underlined.}
\label{overall}
\begin{tabular}{m{0.8cm}|c|cc|cc|cc|cc|cc|cc|cccccc}
\hline
 Victim/ & \multirow{2}{*}{Metric} & \multicolumn{2}{c|}{Pure} & \multicolumn{2}{c|}{Random} & \multicolumn{2}{c|}{Bandwagon} & \multicolumn{2}{c|}{GTA} & \multicolumn{2}{c|}{CLeaR} & \multicolumn{2}{c|}{SSL-attack} & \multicolumn{2}{c}{\textbf{DDSP-G}} & \multicolumn{2}{c}{\textbf{DDSP-B}} \\ 
 Dataset &  & Rec & \cellcolor[HTML]{EFEFEF}Atk & Rec & \cellcolor[HTML]{EFEFEF}Atk & Rec & \cellcolor[HTML]{EFEFEF}Atk & Rec & \cellcolor[HTML]{EFEFEF}Atk & Rec & \cellcolor[HTML]{EFEFEF}Atk & Rec & \cellcolor[HTML]{EFEFEF}Atk & Rec & \cellcolor[HTML]{EFEFEF}Atk & Rec & \cellcolor[HTML]{EFEFEF}Atk\\ \hline
 & H@10 & \textbf{0.0770} & \cellcolor[HTML]{EFEFEF}0.0001 & 0.0730 & \cellcolor[HTML]{EFEFEF}0.0001 & 0.0742 & \cellcolor[HTML]{EFEFEF}0.0003 & 0.0724 & \cellcolor[HTML]{EFEFEF}0.0047 & 0.0727 & \cellcolor[HTML]{EFEFEF}0.0351 & 0.0711 & \cellcolor[HTML]{EFEFEF}0.0245 & \underline{0.0744} & \cellcolor[HTML]{EFEFEF} \underline{0.0404} & 0.0732 & \cellcolor[HTML]{EFEFEF}\textbf{0.0429} \\
 & N@10 & \textbf{0.0422} & \cellcolor[HTML]{EFEFEF}- & 0.0389 & \cellcolor[HTML]{EFEFEF}- & 0.0395 & \cellcolor[HTML]{EFEFEF}0.0001 & 0.0385 & \cellcolor[HTML]{EFEFEF}0.0020 & 0.0388 & \cellcolor[HTML]{EFEFEF}0.0193 & 0.0382 & \cellcolor[HTML]{EFEFEF}0.0109 & \underline{0.0396} & \cellcolor[HTML]{EFEFEF} \underline{0.0221} & \underline{0.0396} & \cellcolor[HTML]{EFEFEF}\textbf{0.0237} \\
 & H@20 & \textbf{0.1098} & \cellcolor[HTML]{EFEFEF}0.0001 & 0.1062 & \cellcolor[HTML]{EFEFEF}0.0002 & \underline{0.1072} & \cellcolor[HTML]{EFEFEF}0.0005 & 0.1043 & \cellcolor[HTML]{EFEFEF}0.0101 & 0.1051 & \cellcolor[HTML]{EFEFEF}0.0546 & 0.1044 & \cellcolor[HTML]{EFEFEF}0.0529 & 0.1066 & \cellcolor[HTML]{EFEFEF}\underline{0.0624} & 0.1064 & \cellcolor[HTML]{EFEFEF}\textbf{0.0667} \\
\multirow{-4}{*}{\begin{tabular}[c]{@{}c@{}} \rotatebox{90}{\parbox{1.0cm}{\centering SASRec\\(Beauty)}} \end{tabular}}
 & N@20 & \textbf{0.0504} & \cellcolor[HTML]{EFEFEF}- & 0.0473 & \cellcolor[HTML]{EFEFEF}- & \underline{0.0478} & \cellcolor[HTML]{EFEFEF}0.0002 & 0.0465 & \cellcolor[HTML]{EFEFEF}0.0034 & 0.0469 & \cellcolor[HTML]{EFEFEF}0.0243 & 0.0466 & \cellcolor[HTML]{EFEFEF}0.0180 & 0.0478 & \cellcolor[HTML]{EFEFEF}\underline{0.0276} & \underline{0.0479} & \cellcolor[HTML]{EFEFEF}\textbf{0.0297}\\ \hline
 & H@10 & \textbf{0.0756} & \cellcolor[HTML]{EFEFEF}- & 0.0727 & \cellcolor[HTML]{EFEFEF}- & 0.0720 & \cellcolor[HTML]{EFEFEF}0.0006 & 0.0741 & \cellcolor[HTML]{EFEFEF}0.0043 & 0.0724 & \cellcolor[HTML]{EFEFEF}0.0034 & 0.0372 & \cellcolor[HTML]{EFEFEF}0.0267 & \underline{0.0742} & \cellcolor[HTML]{EFEFEF}\underline{0.0303} & 0.0733 & \cellcolor[HTML]{EFEFEF}\textbf{0.0317}\\
 & N@10 & \textbf{0.0402} & \cellcolor[HTML]{EFEFEF}- & \underline{0.0400} & \cellcolor[HTML]{EFEFEF}- & 0.0391 & \cellcolor[HTML]{EFEFEF}0.0003 & 0.0390 & \cellcolor[HTML]{EFEFEF}0.0021 & 0.0384 & \cellcolor[HTML]{EFEFEF}0.0016 & 0.0188 & \cellcolor[HTML]{EFEFEF}0.0124 & 0.0394 & \cellcolor[HTML]{EFEFEF}\underline{0.0166}  & 0.0395 & \cellcolor[HTML]{EFEFEF}\textbf{0.0189}\\
 & H@20 & \textbf{0.1104} & \cellcolor[HTML]{EFEFEF}- & 0.1076 & \cellcolor[HTML]{EFEFEF}- & \underline{0.1078} & \cellcolor[HTML]{EFEFEF}0.0007 & 0.1076 & \cellcolor[HTML]{EFEFEF}0.0086 & 0.1064 & \cellcolor[HTML]{EFEFEF}0.0071 & 0.0558 & \cellcolor[HTML]{EFEFEF}0.0446 & 0.1060 & \cellcolor[HTML]{EFEFEF}\textbf{0.0481}  & 0.1064 & \cellcolor[HTML]{EFEFEF}\underline{0.0475}\\
\multirow{-4}{*}{\begin{tabular}[c]{@{}c@{}}\rotatebox{90}{\parbox{1.0cm}{\centering CL4SRec\\(Beauty)}}\end{tabular}} & N@20 & \textbf{0.0487} & \cellcolor[HTML]{EFEFEF}- & \underline{0.0482} & \cellcolor[HTML]{EFEFEF}- & 0.0481 & \cellcolor[HTML]{EFEFEF}0.0003 & 0.0475 & \cellcolor[HTML]{EFEFEF}0.0031 & 0.0470 & \cellcolor[HTML]{EFEFEF}0.0026 & 0.0234 & \cellcolor[HTML]{EFEFEF}0.0169 & 0.0477 & \cellcolor[HTML]{EFEFEF}\underline{0.0211}  & 0.0478 & \cellcolor[HTML]{EFEFEF}\textbf{0.0229}\\ \hline
 & H@10 &\textbf{0.0368}& \cellcolor[HTML]{EFEFEF}0.0003 & 0.0350 & \cellcolor[HTML]{EFEFEF}0.0004 & 0.0357 & \cellcolor[HTML]{EFEFEF}0.0004 & 0.0342 & \cellcolor[HTML]{EFEFEF}0.0087 & 0.0348 & \cellcolor[HTML]{EFEFEF}0.0298 & 0.0359 & \cellcolor[HTML]{EFEFEF}0.0287 & \underline{0.0363} & \cellcolor[HTML]{EFEFEF}\underline{0.0546} & 0.0355 & \cellcolor[HTML]{EFEFEF}\textbf{0.0594} \\
 & N@10 & \textbf{0.0193} & \cellcolor[HTML]{EFEFEF}0.0003 & 0.0178 & \cellcolor[HTML]{EFEFEF}0.0002 & 0.0189 & \cellcolor[HTML]{EFEFEF}0.0002 & 0.0178 & \cellcolor[HTML]{EFEFEF}0.0047 & 0.0181 & \cellcolor[HTML]{EFEFEF}0.0144 & 0.0189 & \cellcolor[HTML]{EFEFEF}0.0126 & \underline{0.0191}& \cellcolor[HTML]{EFEFEF}\underline{0.0302}  & 0.0186 & \cellcolor[HTML]{EFEFEF}\textbf{0.0338}\\
 & H@20 & \textbf{0.0559} & \cellcolor[HTML]{EFEFEF}0.0003 & 0.0550 & \cellcolor[HTML]{EFEFEF}0.0006 & 0.0555 & \cellcolor[HTML]{EFEFEF}0.0013 & 0.0528 & \cellcolor[HTML]{EFEFEF}0.0141 & 0.0548 & \cellcolor[HTML]{EFEFEF}0.0544 &  0.0555 & \cellcolor[HTML]{EFEFEF}0.0623 & \underline{0.0558} & \cellcolor[HTML]{EFEFEF}\underline{0.0830}  & 0.0554 & \cellcolor[HTML]{EFEFEF}\textbf{0.0878}\\
\multirow{-4}{*}{\begin{tabular}[c]{@{}c@{}}\rotatebox{90}{\parbox{1.0cm}{\centering SASRec\\(Sports)}}\end{tabular}} & N@20 & \textbf{0.0242} & \cellcolor[HTML]{EFEFEF}0.0003 & 0.0228 & \cellcolor[HTML]{EFEFEF}0.0002 & 0.0238 & \cellcolor[HTML]{EFEFEF}0.0004 & 0.0225 & \cellcolor[HTML]{EFEFEF}0.0060 & 0.0231 & \cellcolor[HTML]{EFEFEF}0.0206 & 0.0238 & \cellcolor[HTML]{EFEFEF}0.0209 & \underline{0.0240} & \cellcolor[HTML]{EFEFEF} \underline{0.0365}  & 0.0236 & \cellcolor[HTML]{EFEFEF}\textbf{0.0410}\\ \hline
 & H@10 & \textbf{0.0369} & \cellcolor[HTML]{EFEFEF}0.0006 & 0.0348 & \cellcolor[HTML]{EFEFEF}0.0002 & 0.0359 & \cellcolor[HTML]{EFEFEF}0.0009 & \underline{0.0369} & \cellcolor[HTML]{EFEFEF}0.0031 & 0.0365 & \cellcolor[HTML]{EFEFEF}0.0185 & 0.0362 & \cellcolor[HTML]{EFEFEF}0.0327 & 0.0368 & \cellcolor[HTML]{EFEFEF}\underline{0.0402}  & 0.0364 & \cellcolor[HTML]{EFEFEF}\textbf{0.0407}\\
 & N@10 & \textbf{0.0194} & \cellcolor[HTML]{EFEFEF}0.0002 & 0.0181 & \cellcolor[HTML]{EFEFEF}- & 0.0189 & \cellcolor[HTML]{EFEFEF}0.0004 & 0.0189 & \cellcolor[HTML]{EFEFEF}0.0020 & 0.0189 & \cellcolor[HTML]{EFEFEF}0.0093 & 0.0190 & \cellcolor[HTML]{EFEFEF}0.0158 & 0.0192 & \cellcolor[HTML]{EFEFEF}\underline{0.0225}  & \underline{0.0193} & \cellcolor[HTML]{EFEFEF}\textbf{0.0238}\\
 & H@20 & \textbf{0.0575} & \cellcolor[HTML]{EFEFEF}0.0013 & 0.0555 & \cellcolor[HTML]{EFEFEF}0.0004 & 0.0559 & \cellcolor[HTML]{EFEFEF}0.0013 & 0.0557 & \cellcolor[HTML]{EFEFEF}0.0045 & 0.0571 & \cellcolor[HTML]{EFEFEF}0.0323 & 0.0566 & \cellcolor[HTML]{EFEFEF}0.0583 & 0.0562 & \cellcolor[HTML]{EFEFEF}\textbf{0.0610}  & \underline{0.0574} & \cellcolor[HTML]{EFEFEF}\underline{0.0592}\\
\multirow{-4}{*}{\begin{tabular}[c]{@{}c@{}}\rotatebox{90}{\parbox{1.0cm}{\centering CL4SRec\\(Sports)}}\end{tabular}} & N@20 & \textbf{0.0246} & \cellcolor[HTML]{EFEFEF}0.0004 & 0.0233 & \cellcolor[HTML]{EFEFEF}0.0001 & 0.0239 & \cellcolor[HTML]{EFEFEF}0.0005 & 0.0238 & \cellcolor[HTML]{EFEFEF}0.0024 & 0.0241 & \cellcolor[HTML]{EFEFEF}0.0128 & 0.0241 & \cellcolor[HTML]{EFEFEF}0.0222 & 0.0241 & \cellcolor[HTML]{EFEFEF}\underline{0.0277} & \underline{0.0244} & \cellcolor[HTML]{EFEFEF}\textbf{0.0285} \\ \hline
 & H@10 & \textbf{0.0833} & \cellcolor[HTML]{EFEFEF}0.0002 & 0.0801 & \cellcolor[HTML]{EFEFEF}0.0007 & 0.0815 & \cellcolor[HTML]{EFEFEF}0.0008 & 0.0817 & \cellcolor[HTML]{EFEFEF}0.0113 & 0.0813 & \cellcolor[HTML]{EFEFEF}0.0226 & \underline{0.0825} & \cellcolor[HTML]{EFEFEF}0.0430 & \underline{0.0825} & \cellcolor[HTML]{EFEFEF} \underline{0.0440}  & 0.0817 & \cellcolor[HTML]{EFEFEF}\textbf{0.0466}\\
 & N@10 & \textbf{0.0465} & \cellcolor[HTML]{EFEFEF}- & 0.0442 & \cellcolor[HTML]{EFEFEF}0.0003 & 0.0459 & \cellcolor[HTML]{EFEFEF}0.0005 & 0.0463 & \cellcolor[HTML]{EFEFEF}0.0052 & 0.0458 & \cellcolor[HTML]{EFEFEF}0.0115 & 0.0461 & \cellcolor[HTML]{EFEFEF}0.0205 & 0.0461 & \cellcolor[HTML]{EFEFEF} \underline{0.0227}  & \underline{0.0464} & \cellcolor[HTML]{EFEFEF}\textbf{0.0260}\\
 & H@20 & \textbf{0.1147} & \cellcolor[HTML]{EFEFEF}0.0005 & 0.1128 & \cellcolor[HTML]{EFEFEF}0.0012 & 0.1133 & \cellcolor[HTML]{EFEFEF}0.0011 & 0.1126 & \cellcolor[HTML]{EFEFEF}0.0206 & 0.1125 & \cellcolor[HTML]{EFEFEF}0.0375 & \underline{0.1144} & \cellcolor[HTML]{EFEFEF}\textbf{0.0805} & 0.1140 & \cellcolor[HTML]{EFEFEF}\underline{0.0730}  & 0.1125 & \cellcolor[HTML]{EFEFEF}\underline{0.0730}\\
\multirow{-4}{*}{\begin{tabular}[c]{@{}c@{}}\rotatebox{90}{\parbox{1.0cm}{\centering SASRec\\(Toys)}}\end{tabular}} & N@20 & \textbf{0.0545} & \cellcolor[HTML]{EFEFEF}0.0002 & 0.0524 & \cellcolor[HTML]{EFEFEF}0.0004 & 0.0538 & \cellcolor[HTML]{EFEFEF}0.0006 & 0.0541 & \cellcolor[HTML]{EFEFEF}0.0075 & 0.0540 & \cellcolor[HTML]{EFEFEF}0.0152 & 0.0541 & \cellcolor[HTML]{EFEFEF}\underline{0.0300} & 0.0541 & \cellcolor[HTML]{EFEFEF}\underline{0.0300}  & \underline{0.0542} & \cellcolor[HTML]{EFEFEF}\textbf{0.0326}\\ \hline
 & H@10 & \textbf{0.0827} & \cellcolor[HTML]{EFEFEF}0.0005 & 0.0798 & \cellcolor[HTML]{EFEFEF}0.0003 & 0.0806 & \cellcolor[HTML]{EFEFEF}0.0003 & 0.0804 & \cellcolor[HTML]{EFEFEF}0.0108 & 0.0797 & \cellcolor[HTML]{EFEFEF}0.0197 & 0.0797 & \cellcolor[HTML]{EFEFEF}\underline{0.0406} & \underline{0.0808} & \cellcolor[HTML]{EFEFEF}0.0402  & 0.0800 & \cellcolor[HTML]{EFEFEF}\textbf{0.0432}\\
 & N@10 & \textbf{0.0456} & \cellcolor[HTML]{EFEFEF}0.0003 & 0.0446 & \cellcolor[HTML]{EFEFEF}0.0001 & 0.0451 & \cellcolor[HTML]{EFEFEF}0.0002 & 0.0436 & \cellcolor[HTML]{EFEFEF}0.0052 & 0.0449 & \cellcolor[HTML]{EFEFEF}0.0114 & 0.0450 & \cellcolor[HTML]{EFEFEF}0.0192 & 0.0452 & \cellcolor[HTML]{EFEFEF}\underline{0.0217}  & \underline{0.0455} & \cellcolor[HTML]{EFEFEF}\textbf{ 0.0241}\\
 & H@20 & \underline{0.1142} & \cellcolor[HTML]{EFEFEF}0.0009 & 0.1110 & \cellcolor[HTML]{EFEFEF}0.0003 & 0.1131 & \cellcolor[HTML]{EFEFEF}0.0004 & \textbf{0.1145} & \cellcolor[HTML]{EFEFEF}0.0199 & 0.1114 & \cellcolor[HTML]{EFEFEF}0.03003 & 0.1117 & \cellcolor[HTML]{EFEFEF}\textbf{0.0727} & 0.1128 & \cellcolor[HTML]{EFEFEF}0.0634  &  0.1107 & \cellcolor[HTML]{EFEFEF} 0.0677\\
\multirow{-4}{*}{\begin{tabular}[c]{@{}c@{}}\rotatebox{90}{\parbox{1.0cm}{\centering CL4SRec\\(Toys)}}\end{tabular}} & N@20 & \textbf{0.0535} & \cellcolor[HTML]{EFEFEF}0.0004 & 0.0524 & \cellcolor[HTML]{EFEFEF}0.0001 & 0.0532 & \cellcolor[HTML]{EFEFEF}0.0002 & 0.0522 & \cellcolor[HTML]{EFEFEF}0.0075 & 0.0529 & \cellcolor[HTML]{EFEFEF}0.0140 & 0.0530 & \cellcolor[HTML]{EFEFEF}0.0272 & \underline{0.0533} & \cellcolor[HTML]{EFEFEF}\underline{0.0275}  &  0.0532 & \cellcolor[HTML]{EFEFEF}\textbf{0.0303}\\ \hline
\end{tabular}
\end{table*}

\begin{table*}[]
\setlength{\abovecaptionskip}{0.0cm}
\setlength{\belowcaptionskip}{0.0cm}
\small
\caption{Ablation study. The effect of dual-promotion objective, contrastive regularizer and diversity-aware sequence generation of DDSP on three datasets. The best performance is shown in bold, and the second-best is underlined.}
\label{ablation}
\setlength{\tabcolsep}{1.8mm}{
\begin{tabular}{c|cl|cl|clclclcl|cccc|cclclcl}
\hline
 & \multicolumn{2}{c|}{} & \multicolumn{2}{c|}{} & \multicolumn{8}{c|}{Beauty} & \multicolumn{4}{c|}{Sports} & \multicolumn{7}{c}{Toys} \\ \cline{6-24} 
\multirow{-2}{*}{{Victim}} & \multicolumn{2}{c|}{\multirow{-2}{*}{{DDSP Variant}}} & \multicolumn{2}{c|}{\multirow{-2}{*}{{Task}}} & \multicolumn{2}{c}{{\color[HTML]{000000} H@10}} & \multicolumn{2}{c}{{\color[HTML]{000000} N@10}} & \multicolumn{2}{c}{{\color[HTML]{000000} H@20}} & \multicolumn{2}{c|}{{\color[HTML]{000000} N@20}} & H@10 & N@10 & H@20 & N@20 & {\color[HTML]{000000} H@10} & \multicolumn{2}{c}{{\color[HTML]{000000} N@10}} & \multicolumn{2}{c}{{\color[HTML]{000000} H@20}} & \multicolumn{2}{c}{{\color[HTML]{000000} N@20}} \\ \hline
 & \multicolumn{2}{c|}{Pure} & \multicolumn{2}{c|}{Rec} & \multicolumn{2}{c}{\textbf{0.0770}} & \multicolumn{2}{c}{\textbf{0.0422}} & \multicolumn{2}{c}{\textbf{0.1098}} & \multicolumn{2}{c|}{\textbf{0.0504}} & \textbf{0.0368} & \textbf{0.0193} & \textbf{0.0559} & \textbf{0.0242} & \textbf{0.0833} & \multicolumn{2}{c}{\textbf{0.0465}} & \multicolumn{2}{c}{\textbf{0.1147}} & \multicolumn{2}{c}{\textbf{0.0545}} \\
\cline{2-24} 
 & \multicolumn{2}{c|}{} & \multicolumn{2}{c|}{Rec} & \multicolumn{2}{c}{0.0738} & \multicolumn{2}{c}{0.0394} & \multicolumn{2}{c}{0.1060} & \multicolumn{2}{c|}{0.0474} & {\ul 0.0367} & {\ul 0.0191} & 0.0567 & 0.0241 & 0.0803 & \multicolumn{2}{c}{0.0447} & \multicolumn{2}{c}{0.1115} & \multicolumn{2}{c}{0.0526} \\
 & \multicolumn{2}{c|}{\multirow{-2}{*}{w/o CL\&DIV}} & \multicolumn{2}{c|}{\cellcolor[HTML]{EFEFEF}Atk} & \multicolumn{2}{c}{\cellcolor[HTML]{EFEFEF}0.0303} & \multicolumn{2}{c}{\cellcolor[HTML]{EFEFEF}0.0167} & \multicolumn{2}{c}{\cellcolor[HTML]{EFEFEF}0.0484} & \multicolumn{2}{c|}{\cellcolor[HTML]{EFEFEF}0.0213} & \cellcolor[HTML]{EFEFEF}0.0348 & \cellcolor[HTML]{EFEFEF}0.0181 & \cellcolor[HTML]{EFEFEF}0.0569 & \cellcolor[HTML]{EFEFEF}0.0236 & \cellcolor[HTML]{EFEFEF}{\ul 0.0406} & \multicolumn{2}{c}{\cellcolor[HTML]{EFEFEF}0.0200} & \multicolumn{2}{c}{\cellcolor[HTML]{EFEFEF}{\ul 0.0723}} & \multicolumn{2}{c}{\cellcolor[HTML]{EFEFEF}{\ul 0.0280}} \\
 & \multicolumn{2}{c|}{} & \multicolumn{2}{c|}{Rec} & \multicolumn{2}{c}{0.0724} & \multicolumn{2}{c}{0.0390} & \multicolumn{2}{c}{0.1055} & \multicolumn{2}{c|}{0.0473} & 0.0364 & {\ul 0.0191} & {\ul 0.0568} & {\ul 0.0242} & 0.0825 & \multicolumn{2}{c}{{\ul 0.0463}} & \multicolumn{2}{c}{{\ul 0.1141}} & \multicolumn{2}{c}{{\ul 0.0542}} \\
 & \multicolumn{2}{c|}{\multirow{-2}{*}{w/o DIV}} & \multicolumn{2}{c|}{\cellcolor[HTML]{EFEFEF}Atk} & \multicolumn{2}{c}{\cellcolor[HTML]{EFEFEF}{\ul 0.0334}} & \multicolumn{2}{c}{\cellcolor[HTML]{EFEFEF}{\ul 0.0180}} & \multicolumn{2}{c}{\cellcolor[HTML]{EFEFEF}{\ul 0.0538}} & \multicolumn{2}{c|}{\cellcolor[HTML]{EFEFEF}{\ul 0.0232}} & \cellcolor[HTML]{EFEFEF}{\ul 0.0461} & \cellcolor[HTML]{EFEFEF}{\ul 0.0249} & \cellcolor[HTML]{EFEFEF}{\ul 0.0714} & \cellcolor[HTML]{EFEFEF}{\ul 0.0313} & \cellcolor[HTML]{EFEFEF}{\textbf{0.0410}} & \multicolumn{2}{c}{\cellcolor[HTML]{EFEFEF}{\ul 0.0205}} & \multicolumn{2}{c}{\cellcolor[HTML]{EFEFEF}0.0706} & \multicolumn{2}{c}{\cellcolor[HTML]{EFEFEF}0.0279} \\
 & \multicolumn{2}{c|}{} & \multicolumn{2}{c|}{Rec} & \multicolumn{2}{c}{\underline{0.0744}} & \multicolumn{2}{c}{\underline{0.0396}} & \multicolumn{2}{c}{\underline{0.1066}} & \multicolumn{2}{c|}{\underline{0.0478}} & 0.0363 & \textbf{0.0191} & 0.0558 & 0.0240 & {\ul 0.0825} & \multicolumn{2}{c}{0.0461} & \multicolumn{2}{c}{0.1140} & \multicolumn{2}{c}{0.0541} \\
\multirow{-8}{*}{SASRec} & \multicolumn{2}{c|}{\multirow{-2}{*}{\textbf{DDSP}}} & \multicolumn{2}{c|}{\cellcolor[HTML]{EFEFEF}Atk} & \multicolumn{2}{c}{\cellcolor[HTML]{EFEFEF}\textbf{0.0404}} & \multicolumn{2}{c}{\cellcolor[HTML]{EFEFEF}\textbf{0.0221}} & \multicolumn{2}{c}{\cellcolor[HTML]{EFEFEF}\textbf{0.0624}} & \multicolumn{2}{c|}{\cellcolor[HTML]{EFEFEF}\textbf{0.0276}} & \cellcolor[HTML]{EFEFEF}\textbf{0.0546} & \cellcolor[HTML]{EFEFEF}\textbf{0.0302} & \cellcolor[HTML]{EFEFEF}\textbf{0.0830} & \cellcolor[HTML]{EFEFEF}\textbf{0.0365} & \cellcolor[HTML]{EFEFEF} 0.0402 & \multicolumn{2}{c}{\cellcolor[HTML]{EFEFEF}\textbf{0.0227}} & \multicolumn{2}{c}{\cellcolor[HTML]{EFEFEF}\textbf{0.0730}} & \multicolumn{2}{c}{\cellcolor[HTML]{EFEFEF}\textbf{0.0300}} \\ \hline
 & \multicolumn{2}{c|}{Pure} & \multicolumn{2}{c|}{Rec} & \multicolumn{2}{c}{\textbf{0.0756}} & \multicolumn{2}{c}{\textbf{0.0402}} & \multicolumn{2}{c}{\textbf{0.1104}} & \multicolumn{2}{c|}{\textbf{0.0487}} & \textbf{0.0369} & \textbf{0.0194} & \textbf{0.0575} & \textbf{0.0246} & \textbf{0.0827} & \multicolumn{2}{c}{\textbf{0.0456}} & \multicolumn{2}{c}{\textbf{0.1142}} & \multicolumn{2}{c}{\textbf{0.0535}} \\
  \cline{2-24} 
 & \multicolumn{2}{c|}{} & \multicolumn{2}{c|}{Rec} & \multicolumn{2}{c}{0.0740} & \multicolumn{2}{c}{0.0399} & \multicolumn{2}{c}{{\ul 0.1079}} & \multicolumn{2}{c|}{{\ul 0.0484}} & {\ul 0.0371} & {\ul 0.0193} & {\ul 0.0568} & {\ul 0.0242} & 0.0807 & \multicolumn{2}{c}{0.0445} & \multicolumn{2}{c}{0.1123} & \multicolumn{2}{c}{0.0524} \\
 & \multicolumn{2}{c|}{\multirow{-2}{*}{w/o CL\&DIV}} & \multicolumn{2}{c|}{\cellcolor[HTML]{EFEFEF}Atk} & \multicolumn{2}{c}{\cellcolor[HTML]{EFEFEF}0.0198} & \multicolumn{2}{c}{\cellcolor[HTML]{EFEFEF}0.0111} & \multicolumn{2}{c}{\cellcolor[HTML]{EFEFEF}0.0312} & \multicolumn{2}{c|}{\cellcolor[HTML]{EFEFEF}0.0140} & \cellcolor[HTML]{EFEFEF}{\ul 0.0332} & \cellcolor[HTML]{EFEFEF}0.0180 & \cellcolor[HTML]{EFEFEF}{\ul 0.0522} & \cellcolor[HTML]{EFEFEF}0.0227 & \cellcolor[HTML]{EFEFEF}{\ul 0.0388} & \multicolumn{2}{c}{\cellcolor[HTML]{EFEFEF}{\ul 0.0203}} & \multicolumn{2}{c}{\cellcolor[HTML]{EFEFEF}{\ul 0.0636}} & \multicolumn{2}{c}{\cellcolor[HTML]{EFEFEF}{\ul 0.0265}} \\
 & \multicolumn{2}{c|}{} & \multicolumn{2}{c|}{Rec} & \multicolumn{2}{c}{{\ul 0.0742}} & \multicolumn{2}{c}{{\ul 0.0401}} & \multicolumn{2}{c}{0.1067} & \multicolumn{2}{c|}{0.0483} & {\ul 0.0371} & {\ul 0.0193} & 0.0564 & {\ul 0.0242} & {\ul 0.0808} & \multicolumn{2}{c}{0.0450} & \multicolumn{2}{c}{0.1126} & \multicolumn{2}{c}{0.0530} \\
 & \multicolumn{2}{c|}{\multirow{-2}{*}{w/o DIV}} & \multicolumn{2}{c|}{\cellcolor[HTML]{EFEFEF}Atk} & \multicolumn{2}{c}{\cellcolor[HTML]{EFEFEF}{\ul 0.0289}} & \multicolumn{2}{c}{\cellcolor[HTML]{EFEFEF}{\ul 0.0160}} & \multicolumn{2}{c}{\cellcolor[HTML]{EFEFEF}{\ul 0.0441}} & \multicolumn{2}{c|}{\cellcolor[HTML]{EFEFEF}{\ul 0.0198}} & \cellcolor[HTML]{EFEFEF}0.0331 & \cellcolor[HTML]{EFEFEF}{\ul 0.0183} & \cellcolor[HTML]{EFEFEF}0.0515 & \cellcolor[HTML]{EFEFEF}{\ul 0.0229} & \cellcolor[HTML]{EFEFEF}0.0377 & \multicolumn{2}{c}{\cellcolor[HTML]{EFEFEF}0.0191} & \multicolumn{2}{c}{\cellcolor[HTML]{EFEFEF}0.0627} & \multicolumn{2}{c}{\cellcolor[HTML]{EFEFEF}0.0254} \\
 & \multicolumn{2}{c|}{} & \multicolumn{2}{c|}{Rec} & \multicolumn{2}{c}{0.0742} & \multicolumn{2}{c}{0.0394} & \multicolumn{2}{c}{0.1060} & \multicolumn{2}{c|}{0.0477} & 0.0368 & 0.0192 & 0.0562 & 0.0239 & {\ul 0.0808} & \multicolumn{2}{c}{{\ul 0.0452}} & \multicolumn{2}{c}{{\ul 0.1128}} & \multicolumn{2}{c}{{\ul 0.0533}} \\
\multirow{-8}{*}{CL4SRec} & \multicolumn{2}{c|}{\multirow{-2}{*}{\textbf{DDSP}}} & \multicolumn{2}{c|}{\cellcolor[HTML]{EFEFEF}Atk} & \multicolumn{2}{c}{\cellcolor[HTML]{EFEFEF}\textbf{0.0303}} & \multicolumn{2}{c}{\cellcolor[HTML]{EFEFEF}\textbf{0.0166}} & \multicolumn{2}{c}{\cellcolor[HTML]{EFEFEF}\textbf{0.0481}} & \multicolumn{2}{c|}{\cellcolor[HTML]{EFEFEF}\textbf{0.0211}} & \cellcolor[HTML]{EFEFEF}\textbf{0.0402} & \cellcolor[HTML]{EFEFEF}\textbf{0.0225} & \cellcolor[HTML]{EFEFEF}\textbf{0.0610} & \cellcolor[HTML]{EFEFEF}\textbf{0.0300} & \cellcolor[HTML]{EFEFEF}\textbf{0.0402} & \multicolumn{2}{c}{\cellcolor[HTML]{EFEFEF}\textbf{0.0217}} & \multicolumn{2}{c}{\cellcolor[HTML]{EFEFEF}\textbf{0.0634}} & \multicolumn{2}{c}{\cellcolor[HTML]{EFEFEF}\textbf{0.0275}} \\ \hline
\end{tabular}}
\end{table*}

\subsection{Overall Performance}
We compare both recommendation and attack performance for DDSP and other baselines on three real-world datasets, evaluated using two sequential recommendation models: SASRec and CL4SRec. The results are shown in Table \ref{overall}.\\
(1) \textbf{Attack performance (shaded columns)}: DDSP achieves the highest score in most cases. Attack metrics for Pure tend to 0 because the target item is typically unpopular and difficult to promote without an attack. Both Random and Bandwagon show limited success. Random simply places the target item in poisoning sequences without creating new interaction patterns, whereas {Bandwagon} pairs the target item with popular items but fails to align with user preferences. {GTA} and {CLeaR} improve on {Random} and {Bandwagon} by capturing more effective interactions; they do not fully account for the temporal relationships among items, limiting their attack performance in SRS. Although {SSL-attack} is a sequential attack method, our method outperforms it in most cases. Two factors contribute to {DDSP}'s superior performance. First, the proposed attack objective is designed to promote the target item more effectively. Second, the diversity-aware sequence generation increases the co-occurrence rate between the target item and various other items, boosting the attack success rate. {DDSP-B} outperforms {DDSP-G} in most cases, demonstrating that beam search identifies better sequences for fake users compared to greedy search. \\
(2) \textbf{Recommendation performance}: Compared with other baselines, {DDSP} achieves the highest attack success rate while maintaining stronger recommendation performance. This advantage arises from its dual-promotion objective, which consistently aligns the target item and user preferred items in the same optimization direction. By contrast, {GTA} and {CLeaR} focus solely on promoting the target item, and {SSL-attack} leverages GAN-based sequence generation. They cannot jointly promote the target item and preserve high-ranking of user preferred items, resulting in suboptimal recommendation performance. In essence, these baselines overlook recommendation performance when attacking, whereas {DDSP} optimizes for both.
\subsection{Ablation Study}
We investigate the contribution of each component (\ref{part1}, \ref{part3}) by omitting each component and comparing the effects in improving performance. Table \ref{ablation} shows the comparative results. The setting w/o CL\&DIV excludes \ref{part1}, whereas w/o DIV omits only \ref{part3}.

From the attack performance (shaded cells), we observe that w/o CL\&DIV yields the weakest attack in most cases, indicating that merely shortening the distance between the target item and top items in the feature space is insufficient. Adding contrastive enhancement (i.e., w/o DIV) improves the attack by tightening constraints on the target item while distancing it from irrelevant items, thus facilitating promotion. {DDSP} outperforms all ablated versions. Its advantage primarily stems from enhanced sequence diversity in the fake interactions, which raises the co-occurrence rate between the target item and a broad range of items, boosting its recommendation likelihood. From the recommendation performance (unshaded cells), we can see that although the attack reduces the recommendation accuracy, the performance loss under DDSP remains within acceptable limits. This demonstrates that our approach effectively increases attack success while still preserving a reasonable level of recommendation quality.

\vspace{-2mm}
\subsection{The Efficacy of Dual-Promotion Attack Objective}
\begin{figure}
    \setlength{\abovecaptionskip}{0.0cm}
    \setlength{\belowcaptionskip}{0.0cm}
    \centering
    \includegraphics[width=0.9\linewidth]{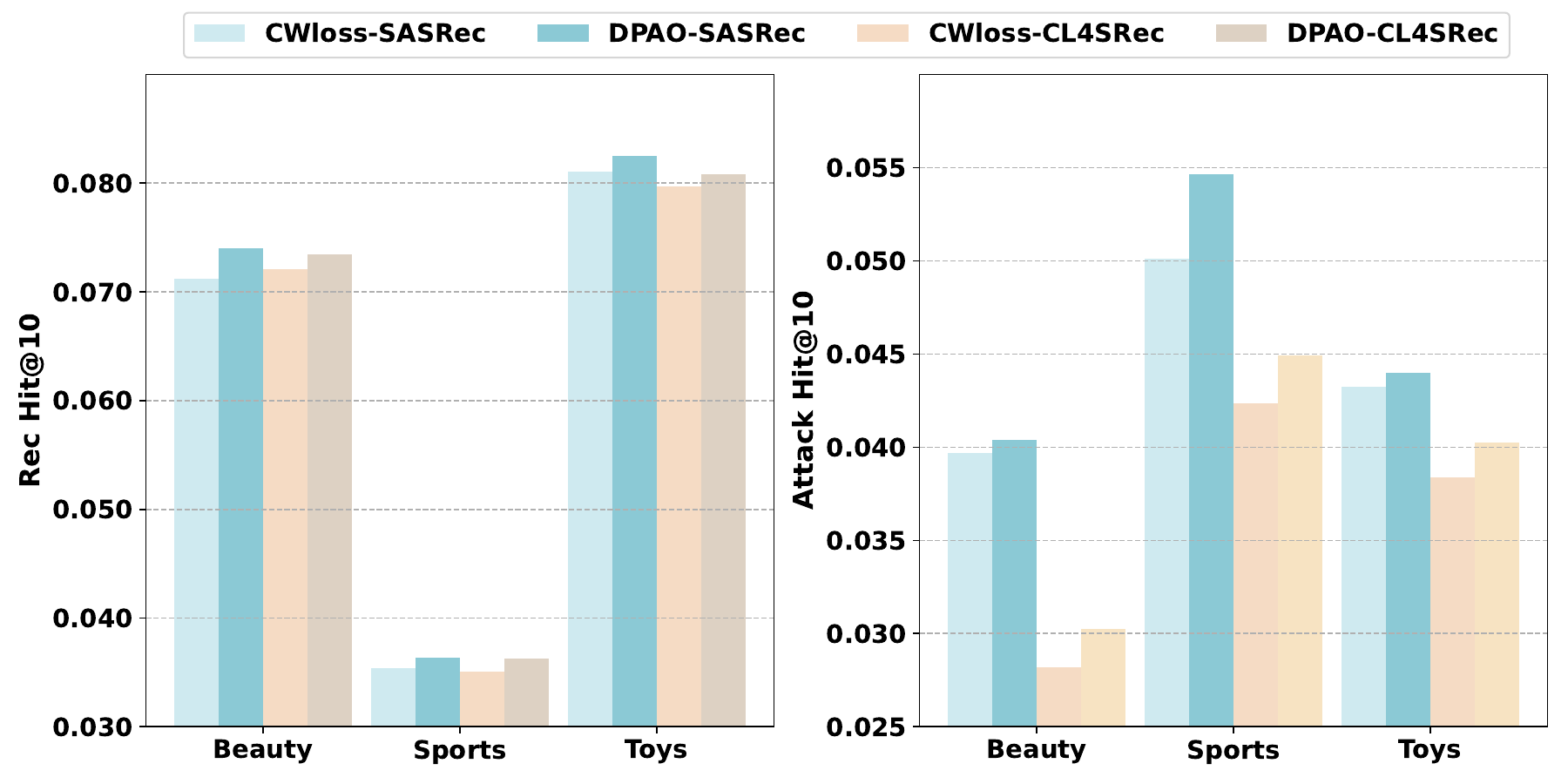}
    \caption{Performance comparison w.r.t. HR@10 of CW loss and DPAO across three datasets on two backbones.}
    \label{lossCompare}
\end{figure}

To evaluate the proposed dual-promotion attack objective ({DPAO}), we replace {DPAO} (Eq.(\ref{l_atk})) with CW loss \cite{rong2022fedrecattack}, a ranking-based attack objective that promotes the target item above the last item in the recommended list. Fig. \ref{lossCompare} presents the results on three datasets under {SASRec} and {CL4SRec}. Experiments show that DPAO consistently outperforms CW loss in both recommendation and attack metrics. The key reason is that {DPAO} naturally aligns with recommendation objectives by boosting the target item alongside user preferred items \cite{li2025htea}. In contrast, CW loss prioritizes raising the target item’s rank at the expense of overall recommendation quality, leading to suboptimal results. Moreover, CW loss only requires the target item to outrank the last item in the list, a weak constraint that diminishes its attack effectiveness.

\vspace{-2mm}
\begin{figure}
    \setlength{\abovecaptionskip}{0.0cm}
    \setlength{\belowcaptionskip}{0.0cm}
    \centering
    \includegraphics[width=\linewidth]{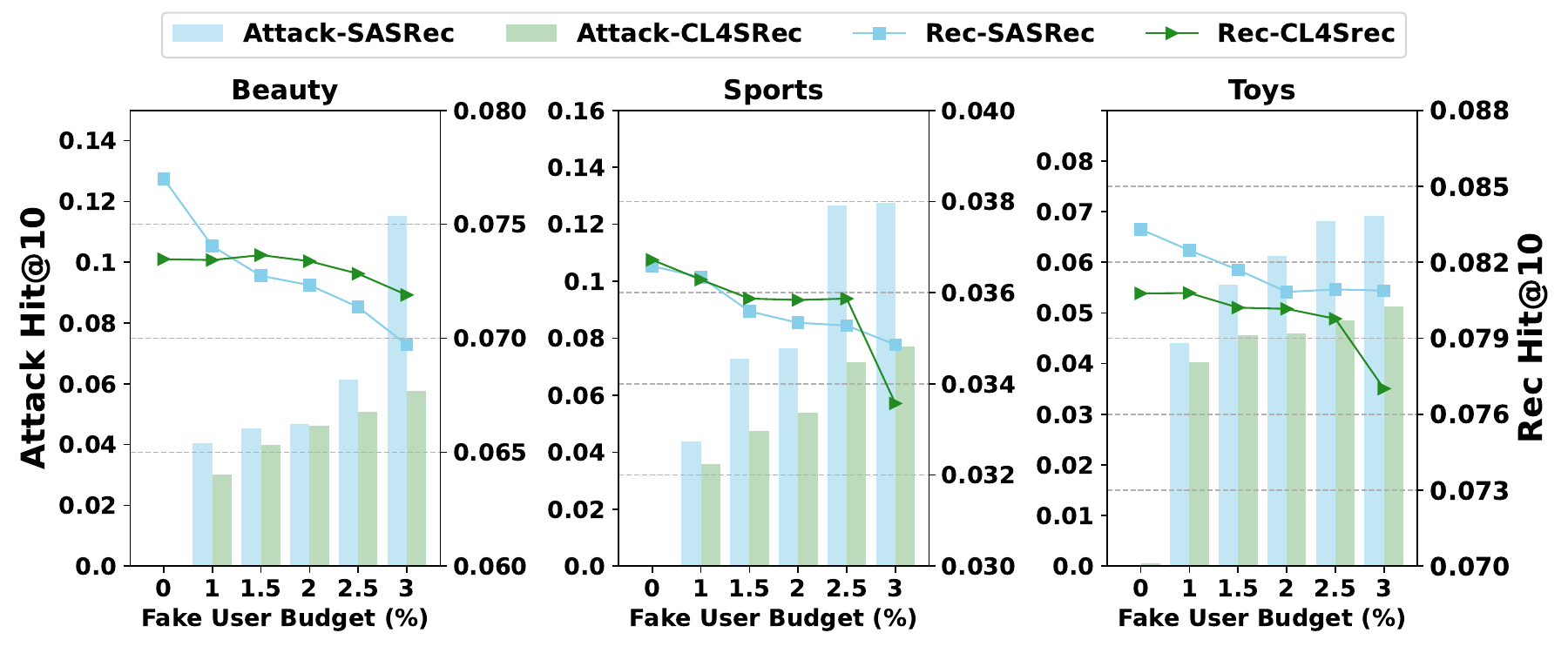}
    \caption{Attack and recommendation performance w.r.t. HR@10 across two datasets under various attack sizes. Bar charts show attack metrics, while line charts indicate recommendation metrics.}
    \label{fakeBudget}
\end{figure}

\subsection{Hyperparameter Analysis}
\subsubsection{\textbf{Impact of Attack Budget}}
We vary the proportion of fake users (1.0\% - 3.0\%) and report the results on three datasets in Fig. \ref{fakeBudget}. As the fake-user budget grows, attack efficacy (bar charts) rises while recommendation performance (line charts) declines. The reason is that having more fake users can boost the attack success rate, but larger perturbations also disrupt the model's learned patterns.
\vspace{-2mm}
\subsubsection{\textbf{Impact of Diversity (\(\lambda\))}}
\begin{figure}
    \setlength{\abovecaptionskip}{0.0cm}
    \setlength{\belowcaptionskip}{0.0cm}
    \centering
    \includegraphics[width=0.95\linewidth]{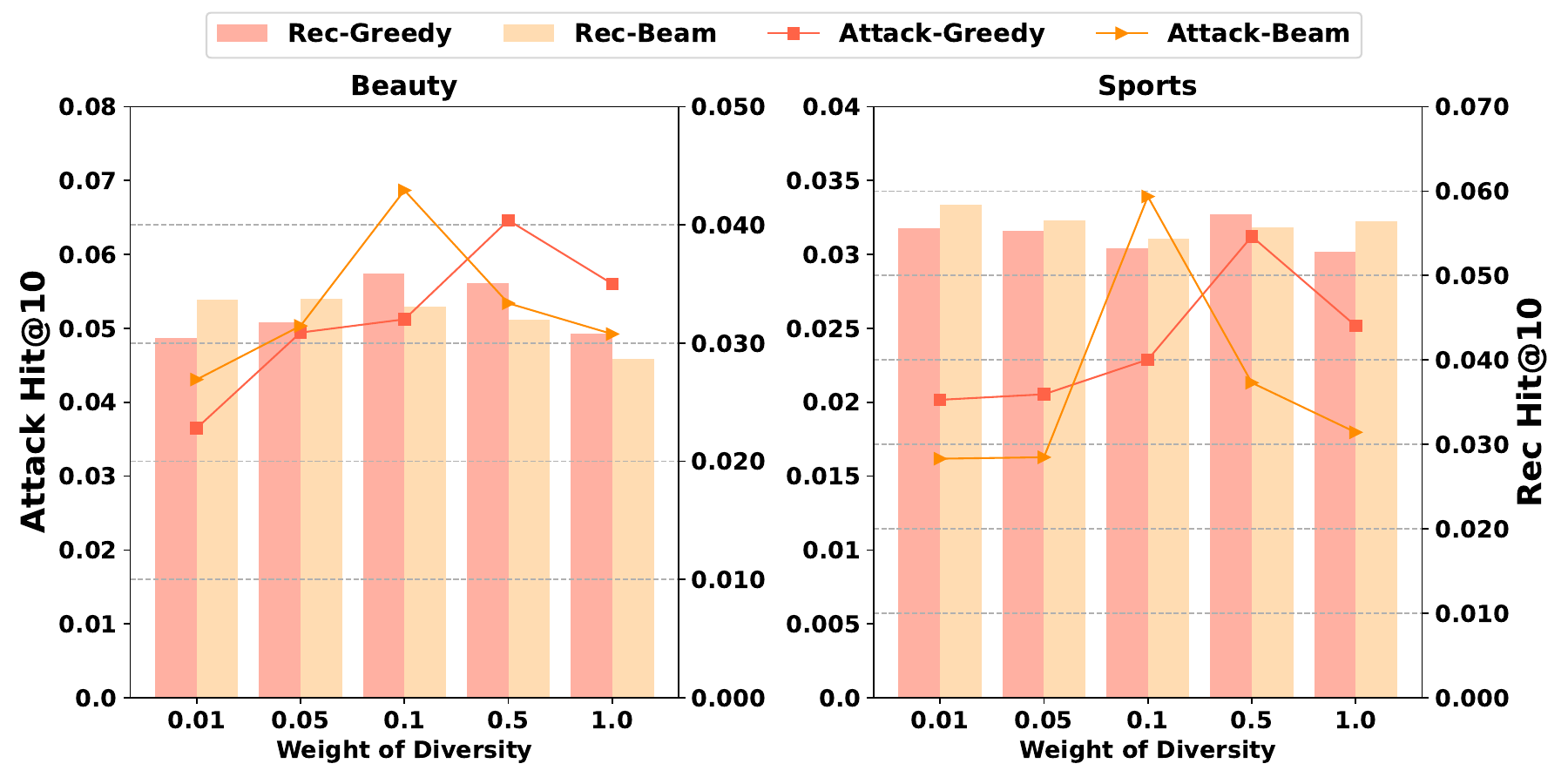}
    \caption{Impact of diversity weight (\(\lambda\)) on 
attack and recommendation quality across two datasets. The bar and line chart illustrate the performance of recommendation and line chart attack metrics based on SASRec.}
    \label{diversity}
\end{figure}

Fig. \ref{diversity} shows how varying \(\lambda\) (0.01 - 1.0) affects both attack success (line charts) and recommendation performance (bar charts). A small \(\lambda\) limits diversity and hinders attack effectiveness. Moderate values strike a balance between diversity and correlation, enhancing the target item promotion. However, excessively large \(\lambda\) disperses item co-occurrences, weakening the model’s ability to associate the target item with relevant items and ultimately harming overall performance.

\vspace{-2mm}
\begin{figure}[h!]
    \centering
    \begin{subfigure}[b]{0.23\textwidth}
        \includegraphics[width=\textwidth]{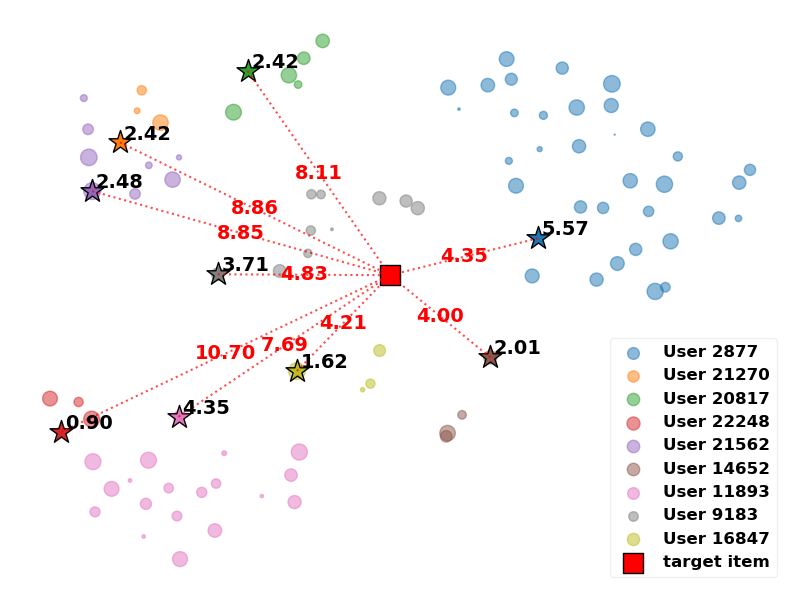}
        \caption{CLeaR}
        \label{clear}
    \end{subfigure}
    \begin{subfigure}[b]{0.23\textwidth}
        \includegraphics[width=\textwidth]{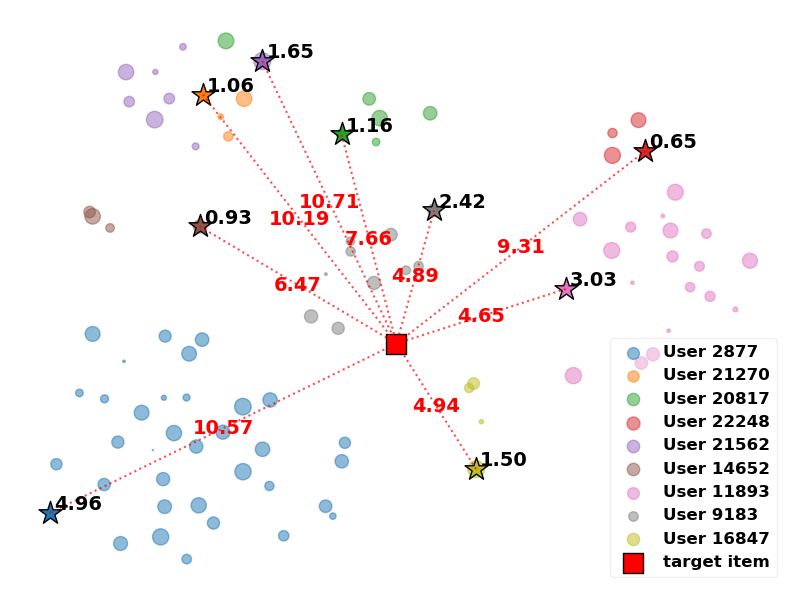}
        \caption{DDSP}
        \label{ddsp}
    \end{subfigure}
    \caption{Visualization on Beauty Dataset of 10 randomly selected users, their ground-truth items, and the target item using t-SNE. The red number indicates each user’s distance to the target item, while the black number next to each user shows their average distance to the ground-truth items.}
    \label{visualization}
\end{figure}

\subsection{Case Study}
We use t-SNE \cite{van2008visualizing} to visualize user and item embeddings on the Beauty dataset, comparing CLeaR and DDSP. Ten users are sampled, along with their ground-truth and target items. In Fig. \ref{visualization}, red labels show distances to the target item, and black labels show distances to ground-truth items. CLeaR places users closer to the target item but farther from their actual preferences, indicating reduced recommendation accuracy. In contrast, DDSP keeps users near both the target item and their ground-truth items, achieving strong attack performance without sacrificing recommendation quality.

\vspace{-2mm}
\section{Conclusion}
In this paper, we conduct a theoretical analysis and identify a conflict between existing recommendation and attack objectives. To address this, we propose a dual-promotion attack objective that simultaneously promotes the target item and user preferred items. Additionally, we design a diversity-aware sequence generation strategy with a re-ranking method to generate fake sequences auto-regressively, enhancing the co-occurrence of the target item with a broader variety of items and improving their authentication. Experiments on three real-world datasets demonstrate that DDSP outperforms existing poisoning attack methods in attack performance and maintains recommendation quality.



\begin{acks}
This work is supported by the Australian Research Council under the streams of Discovery Project (No. DP240101814 and DP240101108), Discovery Early Career Research Award (No. DE230101033 and DE250100613), Future Fellowship (No. FT210100624), and Linkage Project (No. LP230200892 and LP240200546).
\end{acks}

\bibliographystyle{ACM-Reference-Format}
\bibliography{sample-base}

\appendix

\end{document}